\documentclass[onefignum,onetabnum]{siamonline171218}
\usepackage{amssymb}
\usepackage{makecell}
\usepackage{tabularx}
\usepackage{graphicx}
\usepackage{multirow}
\usepackage{float}
\usepackage[algo2e,boxed]{algorithm2e}
\usepackage{subcaption}
\DeclareMathOperator*{\argmin}{arg\,min}
\usepackage[normalem]{ulem}

\newcommand{\bfd}{\mathbf{d}}

\newcommand{\bfr}{\mathbf{r}}
\newcommand{\bfs}{\mathbf{s}}
\newcommand{\bfm}{\mathbf{m}}
\newcommand{\bfu}{\mathbf{u}}

\newcommand{\bfJ}{\mathbf{J}}

\newcommand{\obs}{{\sf {_{obs}}}}
\newcommand{\true}{{\sf {_{true}}}}
\newcommand{\dip}{{\sf {_{dip}}}}
\newcommand{\sfs}{{\sf {_{sfs}}}}
\newcommand{\fwd}{{\sf {_{fwd}}}}
\newcommand{\bwd}{{\sf {_{bwd}}}}

\newcommand{\joint}{{\sf {_{joint}}}}
\newcommand{\ext}{{\sf {_{ext}}}}

\begin{document}

\title{Multi-modal 3D Shape Reconstruction Under Calibration Uncertainty using Parametric Level Set Methods}

\author{Moshe Eliasof\thanks{Computer Science Department, Ben-Gurion University of the Negev, Beer Sheva, Israel. (\email{eliasof@post.bgu.ac.il, asharf@cs.bgu.ac.il, erant@cs,bgu.ac.il})} \and Andrei Sharf\footnotemark[1] \and Eran Treister\footnotemark[1]}

\headers{Shape Reconstruction Under Calibration Uncertainty}{Eliasof M., Sharf A., and Treister E.}

\maketitle

\begin{abstract}
We consider the problem of 3D shape reconstruction from multi-modal data, given uncertain calibration parameters. Typically, 3D data modalities can come in diverse forms such as sparse point sets, volumetric slices, 2D photos and so on. To jointly process these data modalities, we exploit a parametric level set method that utilizes ellipsoidal radial basis functions. This method not only allows us to analytically and compactly represent the object,  it also confers on us the ability to overcome calibration-related noise that originates from inaccurate acquisition parameters. This essentially implicit regularization leads to a highly robust and scalable reconstruction, surpassing other traditional methods. In our results we first demonstrate the ability of the method to compactly represent complex objects. We then show that our reconstruction method is robust both to a small number of measurements and to noise in the acquisition parameters. Finally, we demonstrate our reconstruction abilities from diverse modalities such as volume slices obtained from liquid displacement (similar to CT scans and XRays), and visual measurements obtained from shape silhouettes as well as point-clouds.
\end{abstract}



\section{Introduction}
The reconstruction of 3D objects is an important task in many disciplines like computer graphics \cite{aberman2017dip,avron2010L1,berger2014state,yin2016full,zheng20174d}, computer vision \cite{kolev2009continuous,cremers2011multiview,kolev2012fast}, geophysics \cite{metivier2017full}, computational biology \cite{singer2011three}, civil engineering, medical applications, architecture, and archaeology, among others. Typically, we have incomplete data measurements of a physical object, and wish to reconstruct the full and continuous object that corresponds to these measurements.
In addition to being incomplete, the measurements are usually noisy, and therefore, the aim of the reconstruction process is to find a model that optimally fits the data. This can be done by solving a computational optimization problem, the aim of which is to fit simulated data to existing measurements. These problems are usually ill-posed and non-convex, and solving them with existing methods is often challenging. In many cases---especially when the available data set is small---we need to include some prior information in the reconstruction problem to guide the optimization toward a more plausible solution.

In this work, we focus on problems of 3D shape reconstruction from multi-modal data. That is,  measurements of objects that originate from several domains, such as volumetric slices \cite{aberman2017dip,bermano2011online,kainz2015fast,bretin2017volume,kim2019three,liu2008surface,zou2015topology}, multiview images and object silhouettes \cite{kolev2009continuous,cremers2011multiview,kolev2012fast}, and point-clouds \cite{berger2017survey,qi2017pointnet++}. Our aim is to define a robust unifying framework that can jointly process such data to obtain a best fitting shape model. In the shape reconstruction problem, we are essentially searching for a piecewise constant binary function in some domain, which indicates whether there is an object or background at each location. A common approach to such piecewise constant reconstructions exploits level set methods \cite{santosa1996level,sethian1999level}, which have been used for shape reconstruction \cite{whitaker1998level} and image segmentation \cite{vese2002multiphase}. A particular version of these methods is the parametric level set (PaLS) method \cite{aghasi2011parametric, aghasi2013sparse}, an approach that has been used for inverse problems, particularly for medical \cite{larusson2013parametric, de2015nonlinear} and geophysical \cite{kadu2017salt,mcmillan20153d} applications. The PaLS method suggests that the level set function is approached analytically rather than by using a computational grid. That is, the shape is analytically parameterized by a linear combination of radial basis functions (RBFs) with limited support, e.g., Gaussian functions. The benefit of using RBFs stems from their ability to compactly yet accurately represent the shape and to adapt it as part of the optimization process. Furthermore, we use compactly supported RBFs. This has a tremendous computational benefit when computing a shape from its parameters, leading to sparse Jacobian matrices, and an efficient optimization process for reconstructing high-resolution objects. This method acts as an implicit regularization towards a reconstruction of an dense object.

In the context of graphical applications, RBFs were previously used for surface reconstruction. The work of \cite{carr2001reconstruction} suggests using RBFs for surface reconstruction from a point cloud, where the centers and radii of the basis functions are assumed to be known, and the optimization is performed only for the linear coefficients of the RBFs. A later work \cite{ohtake2003multi} proposed a multi-scale reconstruction that is obtained with RBFs to capture both the global and local structures of the object, after which the reconstructed shape is obtained by interpolating between the scales.
However, such approaches require a large number of basis functions for the reconstruction, in the tens of thousands, conferring a high computational cost on the method. A recent work addressed this issue by devising an efficient GPU implementation of RBF interpolation \cite{cuomo2013surface}. As we will show later, our method requires many fewer RBFs (only a few hundred parameters) to represent whole objects, and not just the surface. A recent work \cite{chen2018deep} proposed employing a deep RBF neural network to encode the spatial distribution of point-clouds for classification. This approach yields competitive classification accuracy (i.e., similar to standard networks \cite{qi2017pointnet++}), while reducing both the number of learned parameters and the computational costs.

In this work we leverage the PaLS method for shape reconstruction, in which the object is analytically represented as a composition of RBFs, and is made piece-wise constant by applying a smoothed step-like function. The parameters of this representation, which is a collection of spheres, include the centers, radii, and linear coefficients of the RBFs. To the best of our knowledge, we are the first group to use PaLS for graphical applications. To that end, our first contribution to this subject features an enriched PaLS representation that comprises a collection of ellipsoids instead of spheres. This is needed since the conventional PaLS method in 3D often results in blob-like artifacts caused by the spherical basis functions. This can be seen in the work of \cite{larusson2013parametric}, which performed 3D medical reconstructions using PaLS. While it is arguable whether such artifacts are acceptable in a medical scenario, they are definitely unacceptable in problems where the reconstructed object should be as close as possible to the original (e.g., reconstruction of a piece of art). To avoid undesired artifacts, therefore, we extend the PaLS representation to be ellipsoidal, so that the PaLS method can compactly represent flat and long objects. As we demonstrate later, our framework is capable of accurately representing complexly shaped objects with sharp elements by using several hundred parameters only. Our approach thus reduces complex reconstruction problems of potentially several million parameters using standard volumetric grid values, to problems of a few thousands parameters \emph{at most} using PaLS to represent the object. The markedly smaller number of parameters using PaLS makes the optimization process efficient and better-posed. As a result, the parametric representation implicitly regularizes the reconstruction and enables us to reconstruct shapes from \emph{small} data sets. Another, albeit smaller benefit of the compact PaLS representation is the savings in storage---we only need to store the PaLS parameters rather than the volumetric grid values of the object.

An additional significant contribution made by our work relates to handling uncertainty in the calibration (or alignment) parameters.
For each measurement, data acquisition methods include certain parameters, e.g., the angle and location relative to the camera at which an image was photographed. The accuracy with which these parameters are known (if they are known at all) varies, generating uncertainty that essentially creates non-trivial noise in the simulated data. Inevitable in real life scenarios, such noise is difficult to model statistically and may hamper the reconstruction if not addressed properly. This is especially evident in cases where the measured data is small. This calibration problem is evident in many reconstruction problems of interest. One may approximate the calibration parameters in addition to the reconstruction, but the process is not straightforward for some problems. For example, in tomography problems there are cases where it is theoretically possible to estimate the calibration (or alignment) parameters and the shape together \cite{basu2000uniqueness}.  Indeed, \cite{houben2011refinement} and \cite{parkinson2012automatic} apply a procedure that alternates between updating the alignment parameters and the tomographic reconstruction. The work of \cite{yang2005unified} optimizes for both the shape and the alignment parameters using a Quasi-Newton method, and the derivatives with respect to the alignment parameters are numerically approximated by finite differences. Similarly, the work of \cite{bleichrodt2013alignment} tackles the problem with the Levenberg-Marquardt algorithm using a spline representation of the shape to treat the derivatives of the alignment parameters. Similarly, \cite{van2018automatic} uses a bi-cubic interpolation of the shape and a variety of first order methods for the joint reconstruction of the shape and alignment parameters. A similar situation is also evident in shape reconstruction from point clouds, which also requires a registration in cases where multiple point clouds are given as data. There, the alignment parameters are usually estimated by the Iterative Closest Point (ICP) methods separately from the reconstruction---see \cite{besl1992method, rusinkiewicz2001efficient, tam2012registration, berger2017survey, tsai2017indoor, takimoto20163d} and references therein. While the current methods are considered to be quite accurate, there may often be small alignment errors, especially if the overlap between the point clouds is small, or if multiple misaligned clouds are considered. In such cases, having additional data sources of the same object can help correcting the alignment. Our solution to this issue is rooted in the PaLS method whose compact analytic representation allows us to apply transformations or deformations to the object. Thus, we can also include the acquisition parameters in the optimization process as part of the multi-modal reconstruction problem, allowing the reconstruction to adapt to the real acquisition parameters, and handle this uncertainty.

We demonstrate both the issue of the accurate object representation, and the handling of uncertainty in the acquisition parameters by performing rigorous experiments using our method for reconstruction from visual and non visual measurements. For the non visual data, we use the acquisition method of ``dip transform'', where the data are obtained by repeatedly dipping the object in liquid, such that each time the object is dipped, it is oriented at a different angle relative to the dipping axis \cite{aberman2017dip}. Using dip transform, though the data acquisition is a simple task, it is tedious to execute, as the data generated from each dip are typically small. To complement this small data, we also experiment with joint shape reconstruction from both the dipping data and visual data, the latter of which comprised silhouette images which are relatively easy to obtain. The visual part of the reconstruction is applied via a modified version of shape from silhouettes (SfS) proposed in \cite{cremers2011multiview,kolev2012fast}. In addition, we perform experiments of shape reconstruction from point clouds.

Our paper is organized as follows: In Section \ref{sec:Formulation}, we provide the mathematical formulations of the problems that we investigate. Then, in Section \ref{sec:palsSection}, we present the PaLS method, after which we address calibration uncertainty in Section \ref{sec:rotParams}. Next in Section \ref{sec:joint}, we discuss the use of two modalities to perform shape reconstruction using the PaLS method. Finally,  in Section \ref{sec:results} we present our empirical results.

\section{Problem Formulation}\label{sec:Formulation}

The typical shape reconstruction problem that we consider here is formulated as a binary piecewise constant optimization problem of the form
\begin{equation}\label{eq:general}
\argmin_{\bfu\in \{0,1\}^{n_1\times n_2\times n_3}} F(\bfu) = \frac{1}{2}\sum_{j=1}^{n_{ex}}\left\|\bfd_j(\bfu) - \bfd_j^{\obs}\right\|_2^2 + \lambda R(\bfu),
\end{equation}
where $n_{ex}$ denotes the number of experiments (e.g., photos), $\bfd_j^{\obs}$ is the data measurement that corresponds to experiment $j$, $\bfu$ is the binary vector representing the object, and $\bfd_j(\bfu)$ is the simulation of the data measurement $j$ according to a given object $\bfu$. We assume that $\bfd_j(\bfu_{\true})\approx \bfd_j^{\obs}$, that is, the simulated data for the true object $\bfu_{\true}$ is approximately equal to the observed data $\bfd_j^{\obs}$. In the first term of $F$ we wish to measure and minimize the simulation error $\bfd(\bfu) - \bfd^{\obs}$ in a least squares sense, which also means that we assume that the simulation error is independent and identically Gaussian distributed sout{and \i.i.d.}. Since the reconstruction problem is often ill-posed, we usually incorporate in it prior information in the form of the regularization term $R(\bfu)$ and its associated parameter $\lambda>0$, the latter of which balances between the data fidelity term and the regularization \cite{vogel2002computational}.

The type of regularization $R(\bfu)$ used can be a simple Tikhonov regularization (as in \cite{aberman2017dip}), or it can be a total variation regularization \cite{rudin1992nonlinear}, the latter of which promotes smoothness in the reconstructed image but is also able to tolerate edges. To solve the problem \eqref{eq:general}, a variety of optimization algorithms can be used, most of which are gradient-based iterative methods such as gradient descent or Gauss-Newton. Such algorithms are suitable only in cases where $\bfu$ is continuous and real-valued. When using these algorithms, therefore, the problem \eqref{eq:general} is solved under the assumption that $\bfu\in\mathbb{R}^{n_1\times n_2\times n_3}$, and $\bfu$ is artificially promoted to be binary by using penalty terms, smoothing regularization and other iterative procedures. This is the case, for example, in \cite{kolev2012fast,kolev2009continuous}, which deal with reconstructions of shapes from silhouettes and multi-view data.
Generally, assigning $\bfu$ to be binary can be a challenge, and may require some heuristics. We demonstrate our framework on the problems below.

\subsection{Shape Reconstruction from Tomographic Measurements}
\label{sec:DipTransform}

\begin{figure}
    \centering
    \begin{subfigure}[b]{0.6\textwidth}
        \includegraphics[width=\textwidth]{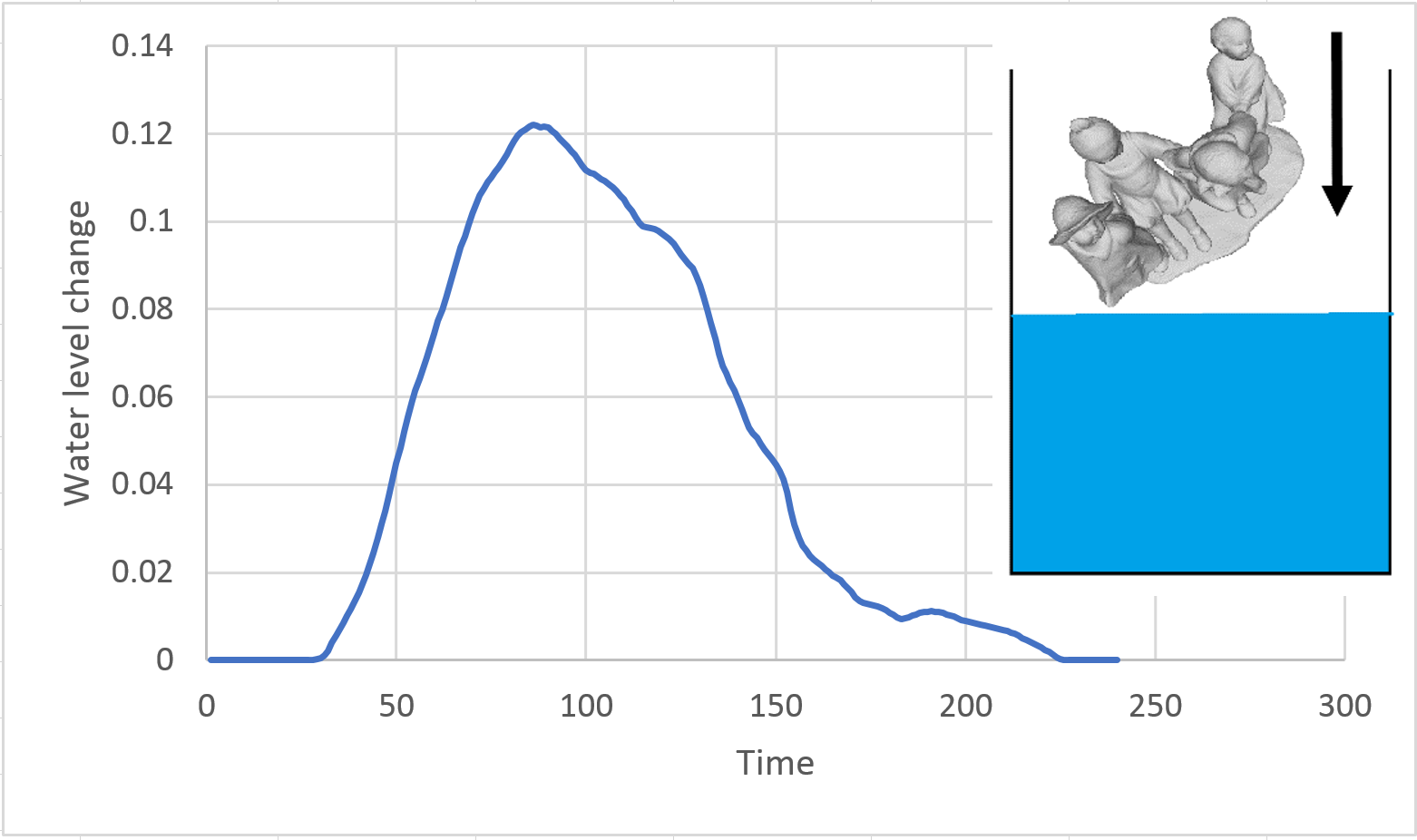}
    \end{subfigure}
\caption{The data acquisition process in the dip transform. An object is repeatedly dipped in water in different orientations. The changes in water level effected by each dip are measured to obtain a signature that describes the object's volume. \vspace{-10pt}}
\label{fig:DipTransform}
\end{figure}

A promising recent tomographic acquisition and shape reconstruction technique is the ``dip transform'' \cite{aberman2017dip}. Designed to reconstruct a shape from liquid displacement measurements that correspond to volumetric slices of the shape, dip transform is similar to other tomographic modalities like X-ray scans, for example, and, in particular, it is not based on visual or surface measurements. Also closely related is the work of \cite{bretin2017volume}, involving reconstruction from 2D slices. The dip transform exploits Archimedes law to reconstruct the volume of the object by repeatedly dipping it in a container full of liquid, such that each time the object is dipped, it is oriented at a different angle relative to a given axis. Each dip along an axis creates a trace of liquid volume displacement that represents the volumes of thin slices of the shape along the dipping axis (see Fig. \ref{fig:DipTransform}). In essence, this technique performs tomographic measurements of the object's slices. Such measurements enable us to reconstruct hidden elements or transparent shapes, which cannot be measured by rays of light.

Using our notation, the dip transform reconstruction problem is defined as:
\begin{equation} \label{eq:dip_original}
\argmin_{\bfu} \left\{ \sum_{j=1}^{n_{dips}}\left\|SP_j\bfu - \bfd_j^{\obs}\right\|_2^2 + \lambda R(\bfu)\right\}, \quad \bfu\in
\{0,1\}^{n_1\times n_2\times n_3}.
\end{equation}
where $P_j$ is a rotation matrix, rotating the object $\bfu$ at horizontal and vertical angles  $\theta_j$ and $\phi_j$, respectively, and essentially is a permutation matrix. $S$ is the ``summation over slices'' matrix, which generates the water displacement measurements given the rotated $\bfu$, and $R$ is the regularization term. Given the angles $(\theta_j,\phi_j)$ and ignoring the binary reconstruction requirement, this is a linear inverse problem.

A significant drawback to this approach is that it necessitates a large number of measurements, as each dip only generates a one-dimensional trace, though we need to reconstruct a three dimensional object. In addition, the mechanical and manual acquisition method begets high uncertainty in the acquisition parameters. In actuality, the angles and centering of the object during dipping are only roughly known. In the next sections, we will show how our method resolves both of these problems.

\subsection{Shape Reconstruction from Surface Measurements}\label{sfs}
\begin{figure}
    \centering
    \begin{subfigure}[b]{0.6\textwidth}
        \includegraphics[width=\textwidth]{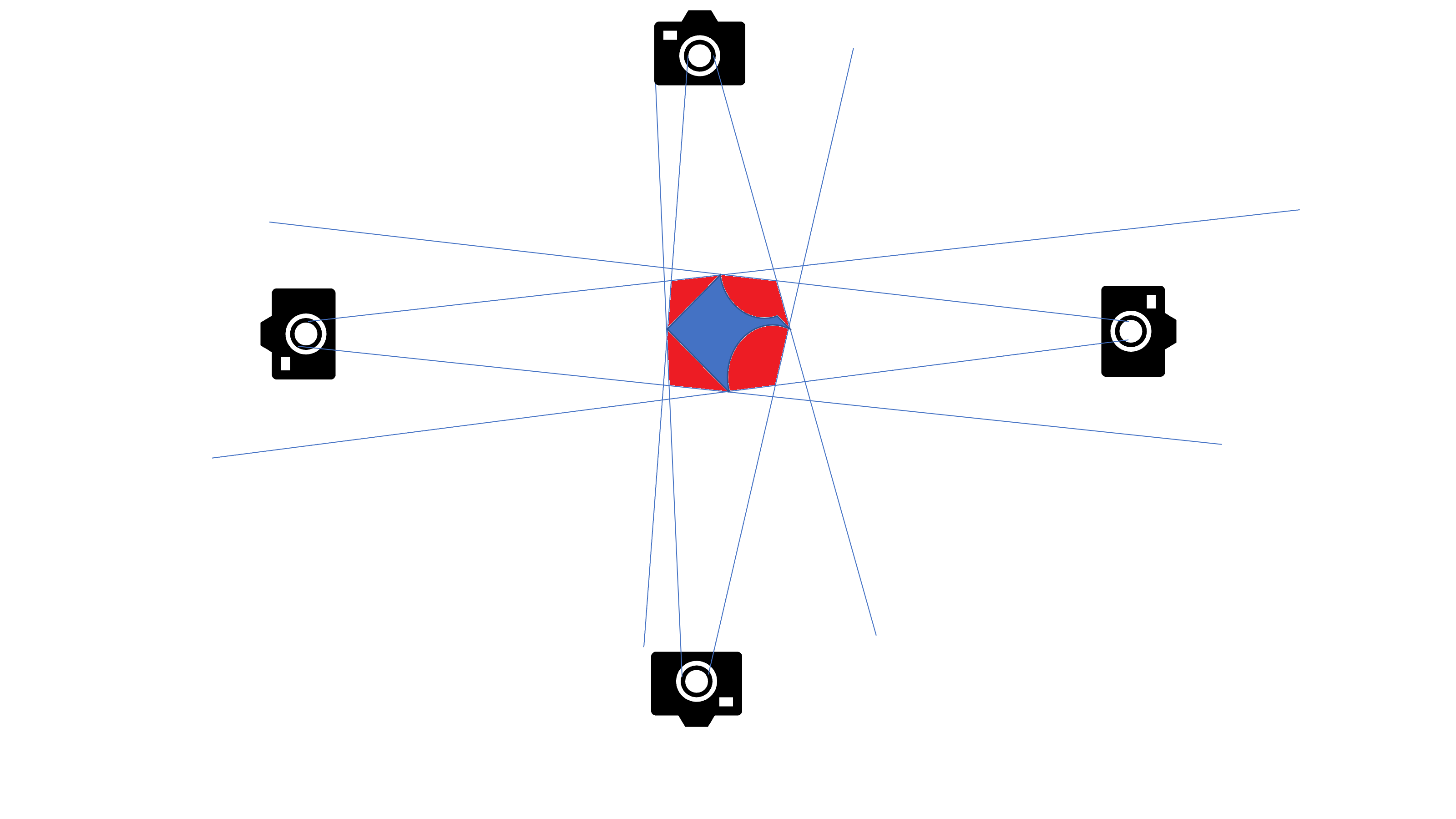}
    \end{subfigure}
\caption{Visual Hull obtained via Shape from Silhouettes, colored in red. The object is colored in blue.}
\label{visualHull}
\end{figure}
Another channel of data input used in this paper comprises visual (or surface) measurements. We consider two data modalities. 
One is the SfS model \cite{laurentini1994visual,kolev2009continuous,cremers2011multiview,kolev2012fast}, whose input includes photographs of the object taken from several different angles. These photographs are then binarized to be shadow-like, and subsequently used to estimate the volumetric shape of the object. Another data modality is point clouds, which are commonly used for object scanning and reconstruction \cite{berger2017survey} using  of-the-shelf sensors. Point clouds consist of sets of points which lie on the (visible) surface of the object. The reconstruction from both modalities may be defined similarly to \eqref{eq:general} as
\begin{equation}
\argmin_{\bfu}\left\{ \sum_j{\mbox{loss}_j(\bfu)} + \lambda R(\bfu)\right\}, \quad \bfu\in \{0,1\}^{n_1\times n_2\times n_3},
\end{equation}
where loss$_j(\bfu)$ combines model principles (either SfS or point cloud) for the $j$-th experiment, and $R$ is again a regularization term. Both modalities are not perfect and have limitations. 
For example, in SfS we cannot satisfactorily reconstruct concavities in the shape, and indeed, in most cases the hidden areas of the shape, like holes or hollows, are artificially filled during the reconstruction process. In point clouds, certain areas of the surface are often scanned in poor resolution, leaving inaccuracies or holes in the reconstructed surface \cite{aberman2017dip}. Nevertheless, both surface models have the advantages of simplicity and ease of the acquisition. 

\subsection{Joint Reconstruction} Each of the data modalities mentioned above has its drawbacks. Therefore, in this work we use multiple modalities to improve the quality of the reconstruction given the available data. For example, we use the SfS to complement the data we acquire via the dip transform, and demonstrate the joint reconstruction of the shape within our framework. To obtain this, we define the joint problem:
\begin{equation}\label{jointMisfitU}
\argmin_{\bfu} \left\{\sum_{j=1}^{n_\dip}\mbox{loss}^{\dip}_j(\bfu) + \gamma \sum_{l=1}^{n_\sfs} \mbox{loss}^{\sfs}_l(\bfu) + \lambda R(\bfu)\right\},
\end{equation}
where $\mbox{loss}^{\dip}$ corresponds to the objective in Eq. \eqref{eq:dip_original}, and $\gamma> 0$ balances the impact of the two loss functions. $\gamma$ is usually chosen such that the two losses are comparable. This combination of SfS with dip transform enables us to achieve a high quality reconstruction using significantly fewer dipping measurements compared to using the dip transform exclusively. In this case, however, we have to neglect the assumption of a full model that is typical for SfS, since the dip measurements are able to capture an object's holes and concavities, which are missed by SfS. Therefore, we propose an adapted SfS model that lacks such assumptions (more details in Section \ref{sec:joint}). In addition, in Section \ref{sec:results} we demonstrate the joint surface reconstruction from SfS and point clouds, to overcome registration problems in point clouds.

\section{Parametric Level Sets using Radial Basis Functions}\label{sec:palsSection}
\begin{figure}
    \centering
    \begin{subfigure}{0.45\textwidth}
        \includegraphics[width=\textwidth,keepaspectratio]{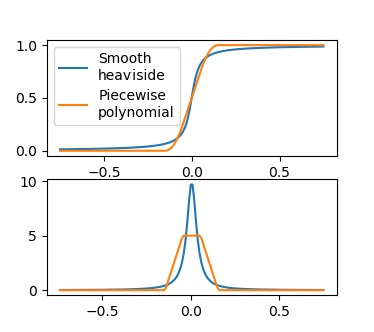}
        \caption{}
        \label{fig:heavyside}
    \end{subfigure}%
    \begin{subfigure}{0.45\textwidth}
        \includegraphics[width=\textwidth,keepaspectratio]{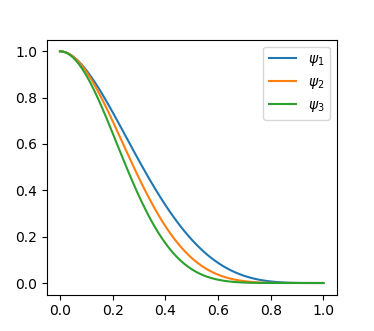}
       \caption{}
       \label{fig:Wendland}
    \end{subfigure}
\caption{The components of the PaLS representation: (a) A smooth vs. piecewise polynomial heaviside function. Top row shows the functions, bottom row shows their derivatives. (b) Wendland's radial basis functions of orders 1,2 and 3.
}
\end{figure}

The PaLS approach \cite{aghasi2011parametric} was originally proposed to solve inverse problems involving a reconstruction of a constant valued body and its background. In our case, in \eqref{eq:general} we have a background of zero, and we focus on retrieving the constant-valued object.
The choice of the basis functions has a significant impact on the dimensionality and expressiveness of the PaLS representation. Herein, we choose radial basis functions \cite{aghasi2011parametric}, which have been shown to represent 3D models well \cite{carr2001reconstruction}. To this end, the PaLS representation is defined by:
\begin{equation}\label{eq:pals}
u(\vec{x},\{\alpha_i,\beta_i,\vec{\xi_i}\}_{i=1}^{n_{RBF}}) = \sigma \left(\sum_i{\alpha_i\psi\left(\left\|\beta_i\left(\vec{x}-\vec{\xi}_i\right)\right\|_2^\dag\right)}\right)
\end{equation}
where $\sigma:\mathbb{R}\rightarrow [0,1]$ is a Heaviside function used to binarize $u$. There are a few functions that can be used for this purpose, like the $\mbox{atan}()$ function used in \cite{aghasi2011parametric}. We leverage a piecewise polynomial Heaviside function as shown in Fig. \ref{fig:heavyside} to improve the condition number of the Hessian and increase the sparsity of the Jacobian \cite{kadu2017parametric}. The function $\psi:\mathbb{R}^+\rightarrow [0,1]$ is one of the Wendland RBF functions that appear in Fig. \ref{fig:Wendland}. For precise definitions of $\psi$ and $\sigma$, refer to the Appendix in Section \ref{sec:heavyside}. The pseudo-norm
$
\left\|\vec{v}\right\|_2^\dag = \sqrt{\left\|\vec{v}\right\|_2^2+\epsilon}
$
is used to prevent division by zero when computing the derivatives for our model.

The RBF together with the $\ell_2$ distance as an operand leads to a sphere in space. Thus, the representation \eqref{eq:pals} is a linear combination of spheres, each one with a radius of approximately $\frac{1}{\beta_{i}}$, centered around $\vec{\xi}_i$. Each basis function has five unknown parameters. We denote:
\begin{equation}\label{eq:m}
\bfm = \{\alpha_i,\beta_i,\vec{\xi_i}\}_{i=1}^{n_{RBF}}\in\mathbb{R}^{5n_{RBF}}
\end{equation}
to be the vector of the unknown RBF parameters in the optimization process, and $\bfu(\bfm)$ to be the discrete version of $u(\vec{x},\bfm)$ which is equivalent to \eqref{eq:pals}.

Next, the problem in Eq. \eqref{eq:general} is replaced with its PaLS version:
\begin{equation}\label{eq:general_PaLS}
\argmin_{\bfm}\hat F(\bfm) =   \sum_{j=1}^{n_{ex}}\left\|\bfd_j(\bfu(\bfm)) - \bfd_j^{obs}\right\|_2^2 + \lambda R(\bfm), \quad \bfm\in
\mathbb{R}^{5n_{RBF}}.
\end{equation}
To solve such a reconstruction problem using a gradient-based method, we need to compute the discrete $\bfu(\bfm)$ and its derivatives with respect to the parameters. That is, we need the Jacobian $\bfJ = \frac{\partial \bfu}{\partial \bfm}$, which we define in Appendix \ref{AppendixPaLS_Derivatives} in detail.
Since our RBFs are compactly supported, the contribution of each basis function is limited by its support, leading to a sparse Jacobian. In addition, each of the derivatives in the Jacobian has a term involving $\sigma'$, which is mostly zero except for the object boundaries, hence the Jacobian matrix $\bfJ$ is very sparse, enabling high resolution reconstructions. While any gradient-based method is suitable to solve a problem like \eqref{eq:general_PaLS}, in this work we use the Gauss-Newton (GN) method, which gives the following iteration step:
\begin{equation}
    \bfm^{(k+1)} = \bfm^{(k)} - \mu^{(k)}(\bfJ^T \bfJ + \lambda \nabla^2 R)^{-1}\nabla_\bfm\hat F,   
\end{equation}
where $\mu^{(k)}$ is obtained by a standard Armijo line-search procedure. For $R$, in this case, we use a simple iterated Tikhonov regularization $R(\bfm) = \|\bfm-\bfm^{(k)}\|_2^2$. This regularization is less sensitive to the choice of $\lambda$ than standard Tikhonov regularization, which does not depend on the iteration $k$ \cite{haber2014computational}. Since $n_{RBF}$---the number of basis functions---is typically small (several hundreds), it is computationally feasible to solve it by using an exact GN, i.e., invert the Hessian $\bfJ^T \bfJ + \lambda \nabla^2 R$ at each iteration. Here, $\lambda$ has to be chosen substantial enough so that the Hessian matrix is numerically inevitable and the optimization is stable. Other suitable options to solve this problem include the non-linear conjugate gradients or LBFGS methods. However, these methods will not be able to fully exploit the low dimension and sparsity of the Hessian, and hence are typically slower in this case.

We start the solution process with a small number of RBFs located at random locations around the center of the grid. Then, at each iteration we add a fixed number of new RBFs to the solution and compute the compact representation of the object on-the-fly. The newly added RBFs are located using the gradient of the misfit function w.r.t $\bfu$ and multiplied by $\sigma'(\bfu)$.
A non-zero $\sigma'(\bfu)$ indicates the areas on the surface of the object where the sensitivities of the PaLS representation are visible (these are the values where $0<\bfu<1$). We also restrict the choice of the new RBF centers to voxels that are distant by at least two grid cells from each other in order not to add basis functions that are too close to each other. We use this heuristic insofar as it is designed to use the least number of parameters possible to reconstruct the object, it guides us to locations in the reconstructed object where its volume is either missing or in excess. The new RBFs are added with the coefficient $\alpha_i=0$, and thus, initially they do not influence the objective of the modeling, resulting in a monotone optimization process. At last, we threshold the solution to make it strictly binary. Algorithm \ref{alg:Reconstruction} summarizes the process.

\begin{algorithm}
\DontPrintSemicolon
\SetKwInOut{Input}{Input}
\SetKwInOut{Init}{Init}
\SetKwInOut{Params}{Params}

\Input{An objective function $F(\bfm)$.}
\Init{Randomly choose $p_0$ basis functions around the center of the grid.}
\Params{$p_0, p$ : The number of the initial and incremental RBFs (default: 20, 5).\\ $it_{GN}$: The number of inner Gauss-Newton iterations (default: 5).
\\ $\delta$: a binarization threshold (default: 0.7). }
 \For{k=1,2,...}{
 \begin{itemize}\Indm
 \item Compute $\bfu = \bfu(\bfm^{(k)})$, and $\nabla_\bfu F$: the gradient of the objective with respect to $\bfu$.
 \item Define the vector $\bfs = |\sigma'(\bfu)\nabla F(\bfu)|$. ($\sigma$: the heaviside function).
 \item Add $p$ new RBFs to $\bfm^{(k)}$, centered at points that correspond to large $\bfs$.
 \item Set up the iterated Tikhonov regularization.
 \item Apply $it_{GN}$ Gauss-Newton iterations for minimizing $\hat F(\bfm)$: $$\bfm^{(k+1)} = \argmin_\bfm\hat F(\bfm),$$
\end{itemize}
}
Binarize $\bfu(\bfm)$ by thresholding: $\bfu = \left\{\begin{array}{lc}1 & if\; \bfu > \delta \\ 0 & otherwise \end{array}\right.$. \;
 \caption{3D Shape Reconstruction using PaLS}
 \label{alg:Reconstruction}
\end{algorithm}

\subsection{Ellipsoidal Radial Basis Functions}\label{sec:enriched}
\begin{figure}
    \centering
    \begin{subfigure}[b]{0.33\textwidth}
        \includegraphics[width=\textwidth,keepaspectratio,trim={5cm 1.3cm 5cm 0cm},clip]{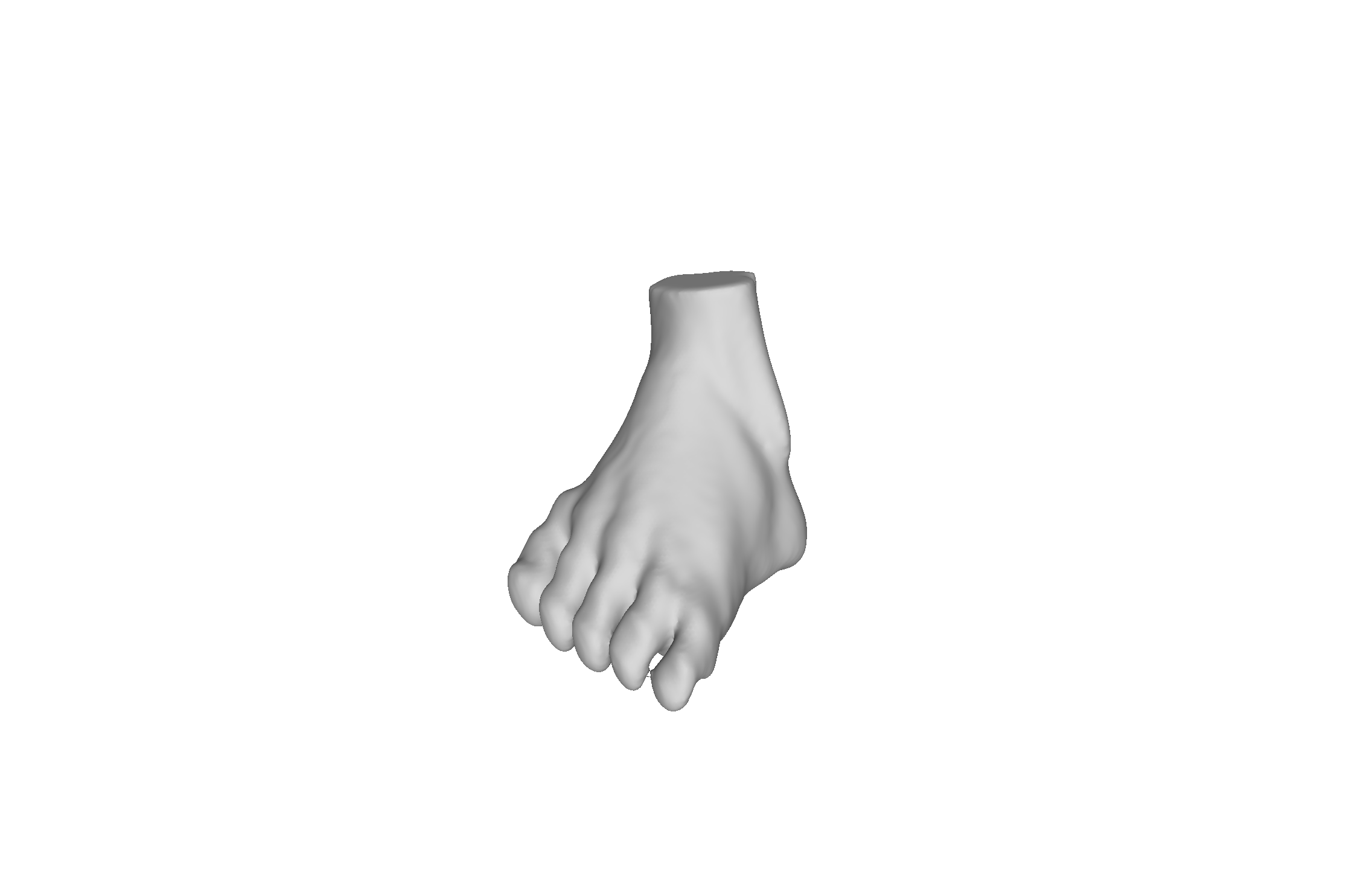}
        \caption{Input}
        \label{fig:feetInput}
    \end{subfigure}%
    \begin{subfigure}[b]{0.33\textwidth}
        \includegraphics[width=\textwidth,keepaspectratio,trim={5cm 2cm 5cm 2cm},clip]{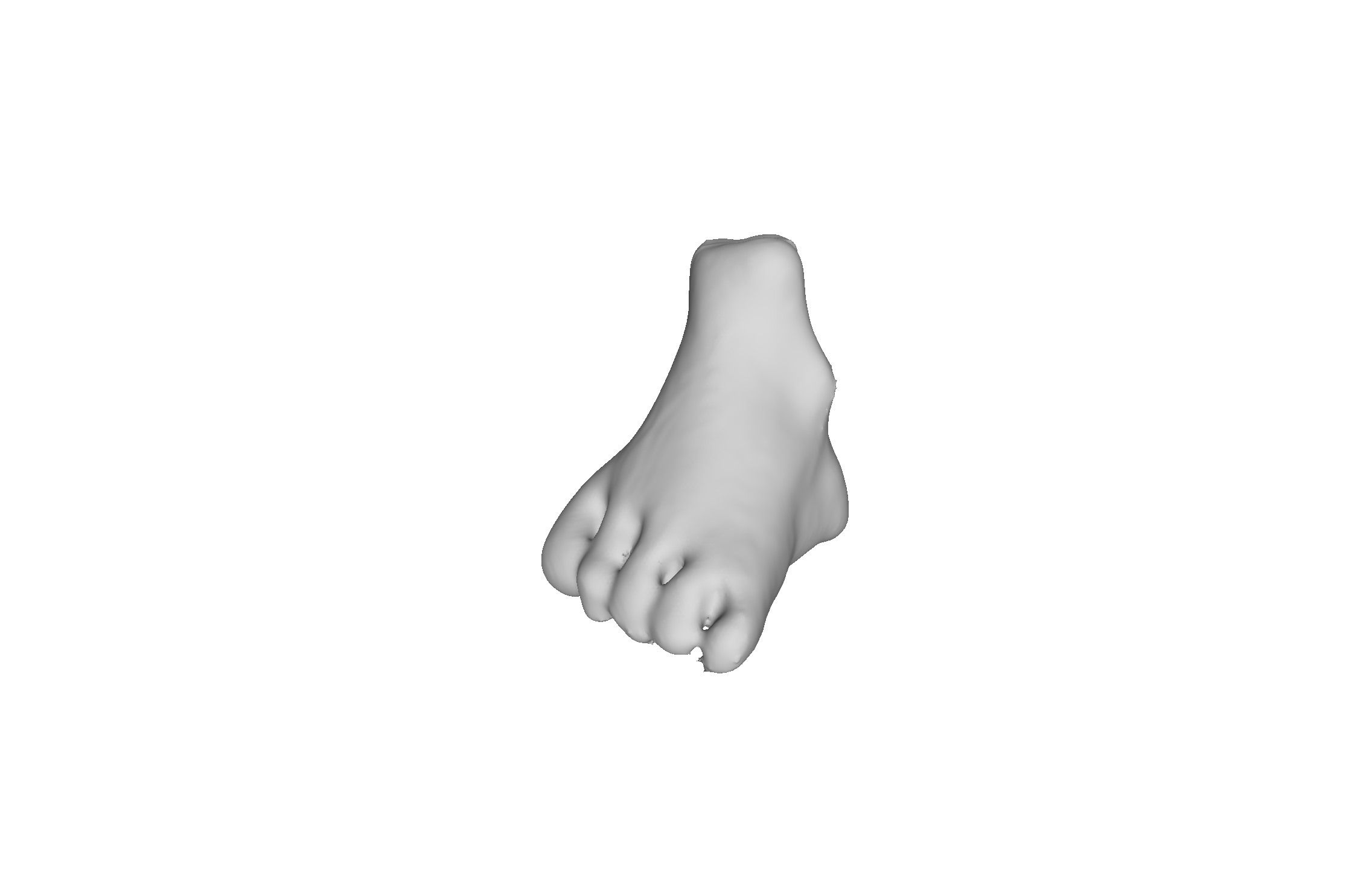}
        \caption{PaLS with spherical RBFs}
        \label{fig:feet5Based}
    \end{subfigure}%
    \begin{subfigure}[b]{0.33\textwidth}
        \includegraphics[width=\textwidth,trim={5cm 2cm 5cm 2cm},clip]{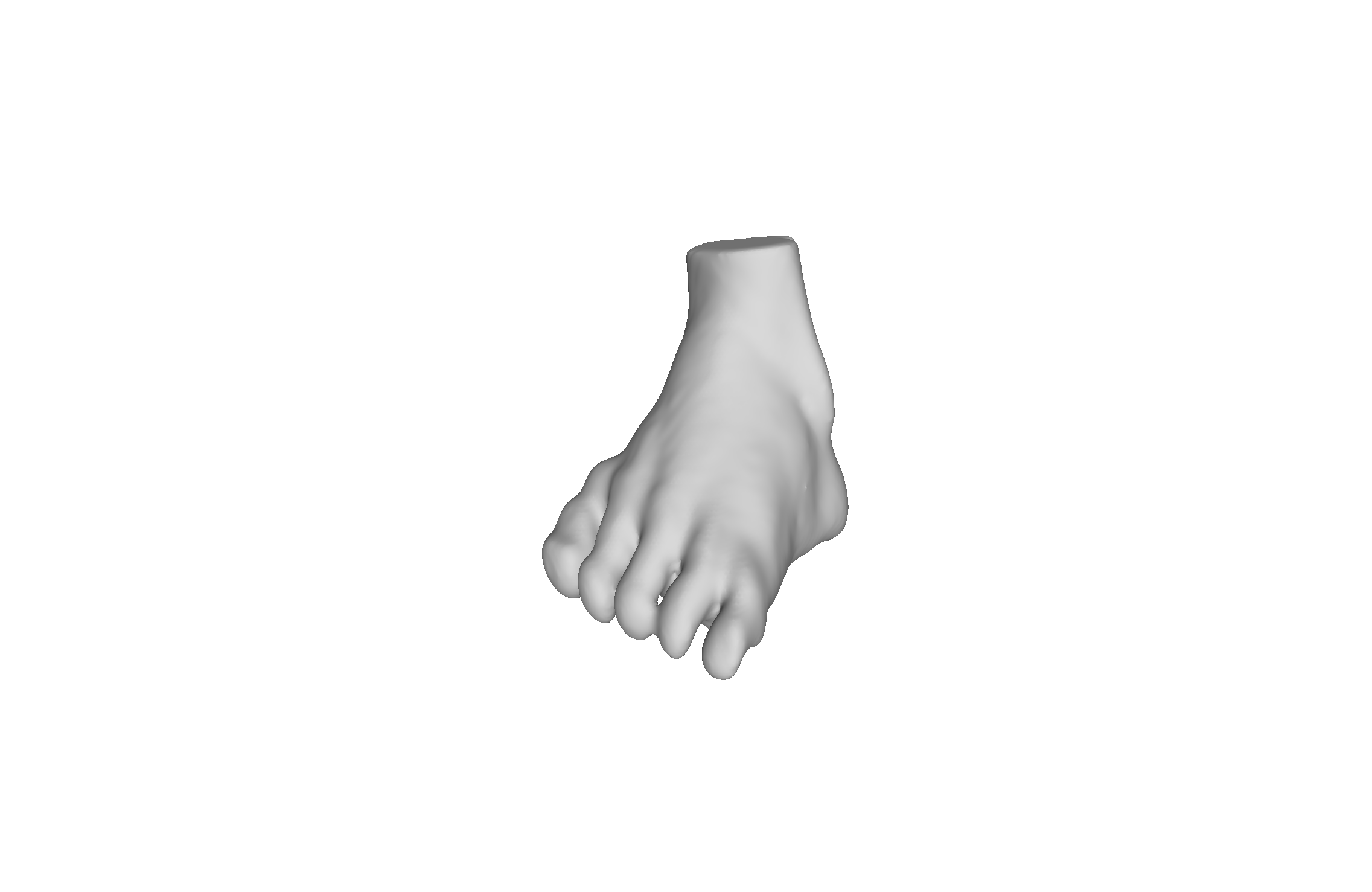}
        \caption{PaLS with ellipsoidal RBFs}
        \label{fig:feet10Based}
    \end{subfigure}
    \caption{Comparison between reconstructions with spherical vs. ellipsoidal PaLS. Both reconstructions use 200 parameters, that is, 40 RBFs for (b) and 20 RBFs for (c). }\label{fig:5vs10}
    \label{fig:Enriched_vs_Basic}
\end{figure}

Using the standard RBF representation in \eqref{eq:pals} restricts us to a spherical basis that results in limited expressiveness and spherical effects in the reconstructed objects, as can be observed in \cite{larusson2013parametric}. Clearly, this outcome is undesirable in computer graphics applications, especially when objects with flat or sharp elements such as arms, legs or horns, are involved.
To render the basis functions more expressive, we suggest that they be enriched to enable them to represent ellipsoids in space rather than spheres. Such enrichment is done by replacing the parameters $\beta_i$ in \eqref{eq:pals} with symmetric and positive definite (SPD) matrices $B_i\in\mathbb{R}^{3\times 3}$. Our ellipsoidal PaLS representation reads:
\begin{equation}\label{enrichedPaLS}
u(\vec{x},\bfm) = \sigma \left(\sum_i{\alpha_i\psi\left(\left\|\left(\vec{x}-\vec{\xi}_i\right)\right\|_{B_i}^\dag\right)}\right),
\end{equation}
where the weighted pseudo norm is given by
$\left\|\vec{v}\right\|_{B_i}^\dag = \sqrt{\left\|\vec{v}\right\|_{B_i}^2+\epsilon}$,
and similar to \eqref{eq:m}, the \emph{ellipsoidal} parameter vector is
\begin{equation}
\bfm = \{\alpha_i,B_i,\vec{\xi_i}\}_{i=1}^{n_{RBF}}\in\mathbb{R}^{10n_{RBF}}.
\end{equation}
Our enriched representation denotes a collection of ellipsoids, each with 10 unknowns. For a description of how the derivatives are computed, see the Appendix in Section \ref{AppendixPaLS_Derivatives}. In this case, we also require a regularization to ensure that the matrices $B_i$ remain SPD. To this end, in addition to the Tikhonov regularization on $\bfm$, we use a standard log barrier function ($-\sum_i\log(\det(B_i))$) to keep the determinants of $B_i$ away from zero. This regularization is effective, as we do not expect $B_i$ to be too close to singular. In other cases where $B_i$ are expected to be close to singular, the more sophisticated alternative approach of optimization over manifolds \cite{boumal2014manopt} may be more suitable.
To demonstrate the advantages of the ellipsoidal over the spherical RBFs, we reconstruct a model of a foot that has both detailed and smooth elements (Fig. \ref{fig:Enriched_vs_Basic}). The figure clearly shows that the proposed ellipsoidal PaLS captures greater detail than the PaLS with basic RBFs, even though the two reconstruction algorithms used the the same number of parameters.

\section{Robustness to Calibration Uncertainty and Scalability}\label{sec:rotParams}

Object reconstruction algorithms typically consume data of the object captured in several orientations, e.g., pictures of the object taken at different angles \cite{cremers2011multiview}.
A major drawback of many algorithms is the assumption that the orientations at which the data were acquired are accurately known and that the noise in the measurements corresponds to a simple probability distribution, e.g., Gaussian. This assumption may be practical in some cases---especially when the data set is large---but it may not be ideal in real-world scenarios with small data.
The absence of accurately known acquisition parameters creates noise in the measurements that is difficult to model or predict. Such noise can also be introduced due to a human factor in the data acquisition process, e.g., hands shaking while capturing the object, or due to mechanical inaccuracies in the machine that rotate the object for data collection as in \cite{aberman2017dip}. This often leads to poor reconstruction, especially in cases where the data set is small. To solve these problems, we use the analytic PaLS representation of the object, which allows us to analytically manipulate the object's position parameters, and we estimate the acquisition parameters as part of \emph{the optimization process for the data fit}. This is possible because we can differentiate the PaLS representation of $\bfu$ with respect to the parameters of the data acquisition.

Consider, for example, the dip transform problem in \eqref{eq:dip_original}.  The PaLS formulation of this problem (analogous to that of the general problem in \eqref{eq:general_PaLS}), is
\begin{equation}\label{eq:nonRotParamEq}
\argmin_{\bfm} \left\{ \sum_{j=1}^{n_{dips}}\left\|SP_j\bfu(\bfm) - \bfd_j^{obs}\right\|_2^2 + \lambda R(\bfm)\right\}.
\end{equation}
Here $P_j$ is a permutation matrix designed to capture the object from different angles and locations that correspond to the $j$-th measurements of the shape, and $S$ is a projection operator that simulates the data after the object has been situated according to the acquisition parameters.
The operator $SP_j$ in \eqref{eq:dip_original} is linear. The matrices $SP_j$, however, may require large amounts of storage or be expensive to apply. More importantly, such fixed operators introduce non trivial noise if the acquisition parameters are not accurately known. Therefore, in addition to $\bfm$, we   also search for the acquisition parameters encapsulated in $P_j$ and define the rotation and translation operator $P_j$ analytically.

To define the rotation and translation in $P_j$ analytically, we first define the linear transformation of our coordinates
\begin{eqnarray}\label{eq:transformation}
T_j(\vec{x}) = Q(\theta_j,\phi_j)(\vec{x} - \vec{x}_{mid}) + \vec{b_j} + \vec{x}_{mid}.
\end{eqnarray}
Here $Q(\theta_j,\phi_j)$ is a $3\times3$ orthogonal rotation matrix in azimuth angle $\theta_j$ and polar angle $\phi_j$.  $\vec{b_j}\in \mathbb{R}^3$ is a translation vector, and $\vec{x}_{mid}$ is the center of the domain that is used as the center of the rotation. This is the transformation that guides the definition of $P_j$ in \eqref{eq:dip_original}. Next, we define the analytically translated and rotated shape $u$ via backward warping according to the transformation \eqref{eq:transformation}:   $u(T_j^{-1}(\vec{x}),\bfm)$.
The backward warping is used to avoid discretization problems in the rotation.
Note, that the rotated $u$ is the analytical definition of the discrete operator $P_j$, which, up to discretization errors on the mesh, satisfy
\begin{equation}\label{eq:rotatedShape}
u(T_j^{-1}(\vec{x}),\bfm) \approx P_j\bfu(\bfm).
\end{equation}
Obviously, the rotated shape can also be represented by PaLS by using what we refer to as ``rotated parameters''
\begin{equation}\label{eq:mrot}
rot_j(\bfm) =rot(\bfm,\theta_j,\phi_j,\vec{b_j}),
\end{equation}
where $rot_j$ is the function for rotating the parameters $\bfm$ in the angles $\theta_j,\phi_j$ and translation vector $\vec{b_j}$, which together correspond to the $j$-th measurements of the shape. We define $rot_j$ below for both versions of the PaLS. To define an inverse problem with analytically rotated parameters (instead of using $P_j$), we can simply replace the term $P\bfu(\bfm)$ in \eqref{eq:nonRotParamEq} with $\bfu(rot_j(\bfm))$. As in \eqref{eq:nonRotParamEq}, this will lead to a reconstruction with fixed acquisition parameters, which we incorporate into the inversion, described in the following section. Next, we explicitly define $rot_j(\bfm)$ for each of the considered formulations, and the definition of their associated Jacobians appears in the Appendix in Section \ref{sec:PaLS_RotJacobians}.

\paragraph{Rotation for Spherical RBFs}
In the simpler version of the PaLS, we plug \eqref{eq:rotatedShape} into \eqref{eq:pals} and obtain
\begin{eqnarray}
u(T^{-1}(\vec{x}),\bfm) &=& \sigma \left(\sum_i{\alpha_i\psi\left(\left\|\beta_i\left(T^{-1}(\vec{x})-\vec{\xi}_i\right)\right\|_2^\dag\right)}\right)\\
&=& \sigma \left(\sum_i{\alpha_i\psi\left(\left\|\beta_i\left(\vec{x}-T_j(\vec{\xi}_i)\right)\right\|_2^\dag\right)}\right)
\end{eqnarray}
The last equality holds because $Q$ is an orthogonal matrix that does not change the $\ell_2$ norm of a vector. Hence, $rot_j(\bfm)$ only rotates the center of each basis function $i$ and is defined by
\begin{equation}\label{eq:simple_mrot}
rot_j\left(\alpha_i,\beta_i,\vec{\xi}_i\right) = rot\left(\alpha_i,\beta_{i},\vec{\xi}_i,\theta_j,\phi_j,\vec{b_j}\right)= \left(\alpha_i,\beta_i,T_j(\vec{\xi}_i)\right)
\end{equation}

\paragraph{Rotation for the Ellipsoidal RBFs}
The rotation with respect to the ellipsoidal RBFs is slightly more complicated than in the spherical case, and is given by
\begin{eqnarray}
u(T^{-1}(\vec{x}),\bfm) &=& \sigma \left(\sum_i{\alpha_i\psi\left(\left\|\left(T^{-1}(\vec{x})-\vec{\xi}_i\right)\right\|_{B_i}^\dag\right)}\right)\\
&=& \sigma \left(\sum_i{\alpha_i\psi\left(\left\|\left( (\vec{x} - \vec{x}_{mid} - \vec{b}_j) + T_{j}(\vec{\xi}_i) \right)\right\|_{Q_j^{-T}B_iQ_j^{-1}}^\dag\right)}\right).
\end{eqnarray}
Here, for the $i$-th basis function defined by $({\alpha_i,B_i,\vec{\xi_i}})$, we denote its rotation and translation according to the $j$-th acquisition parameters as
\begin{equation}\label{eq:rot_enriched}
rot_j\left(\alpha_i,B_{i},\vec{\xi}_i\right) = \left(\alpha_i,Q_{j}^{-T} B_{i} Q_{j}^{-1},T_j(\vec{\xi}_i)\right) = (\alpha_i,Q_{j} B_{i} Q_{j}^{T},T_j(\vec{\xi}_i)),
\end{equation}
where the last equality holds since $Q$ is orthogonal.

\subsection{Handling Calibration Uncertainty in the Data Acquisition}\label{sec:noiseHandling}
So far the acquisition parameters were kept fixed, and we did not address the non-trivial noise that originates from inaccuracies in those parameters. Adequately handling such noise is important to produce an accurate reconstruction. Therefore, as part of the reconstruction we also manipulate the PaLS representation parameters of $\bfu$ with respect to the parameters of the data acquisition $\{(\theta_j , \phi_j , \vec{b}_j)\}_{j=1}^{n_{ex}}$.

To include the data acquisition parameters in the reconstruction, we first extend our (ellipsoidal) inversion parameters:
\begin{equation}\label{eq:mExtendedEnriched}
 \bfm_{\ext}= \left[ {(\alpha_i,B_i, \xi_i)}_{i=1}^{n_{RBF}}, {(\theta_j , \phi_j , \vec{b}_j )}_{j=1}^{n_{ex}} \right] \in\mathbb{R}^{10n_{RBF} + 5n_{ex}}.
\end{equation}
The treatment with respect to the spherical RBFs is similar. To have the parameters of the $i$-th basis function rotated by the \emph{predicted} $j$-th data acquisition parameters, we use the function $rot()$ in \eqref{eq:rot_enriched}, but now let the acquisition parameters change throughout the optimization process.
By combining Eqs.  \eqref{eq:nonRotParamEq}-\eqref{eq:mrot} together, we obtain our final inverse problem
\begin{equation}\label{eq:invProblem_mrot}
\argmin_{\bfm_{\ext}} \left\{ \sum_{j=1}^{n_{ex}}\left\|S\bfu\left(rot\left(\bfm,\left(\theta_j,\phi_j,b_j\right)\right)\right) - \bfd_j^{obs}\right\|_2^2 + \lambda R(\bfm_{\ext})\right\}.
\end{equation}
This is a version of \eqref{eq:nonRotParamEq}, which applies the rotation of the object by rotating its RBF parameters instead of using the matrices $P_j$, and includes the data acquisition parameters as unknowns for the optimization. Similar to the previous cases, we use a Tikhonov regularization for the parameters in $\bfm_{\ext}$.

To solve problem \eqref{eq:invProblem_mrot}, we calculate the sensitivities of the rotated $\bfm$ with respect to up-to-date predictions of $(\theta_j , \phi_j , \vec{b}_j )$ and $(\alpha_i,B_i, \vec{\xi_i})$.
To that end, we define the Jacobian of \eqref{eq:rot_enriched} with respect to all of its parameters:
\begin{equation}\label{eq:JacobianEq}
 \bfJ^{rot} =  \frac{\partial (\alpha_i,Q_{j} B_{i} Q_{j}^{T},T_j(\vec{\xi}_i))}{\partial(\theta_j , \phi_j , \vec{b}_j,\alpha_i,B_i, \vec{\xi_i} )   }.
 \end{equation}
This Jacobian is defined for all pairs of basis functions and acquisition parameters, and using the chain rule it is multiplies with the PaLS Jacobian $\frac{\partial\bfu}{\partial\bfm}$ to capture the sensitivity of $\bfu$ with respect to all the parameters. All the definitions of $\bfJ^{rot}$ are found in Appendix \ref{appendixNoiseHandling}.

\section{Joint Reconstruction using PaLS under Calibration Uncertainty}\label{sec:joint}
In sections \ref{sec:DipTransform}-\ref{sfs}, we present two models that are used to reconstruct a three-dimensional object.
One of the models, the dip transform, includes information on the volume of the object, but it is also characterized by a complex data acquisition mechanism. The second model, SfS, can only provide information about the object's surface (as it is visual), but the data are cheap and easy to acquire. Here we demonstrate how, in our framework, these two models can complement one another. Specifically, using the SfS model together with the dip transform can significantly reduce the number of dips required to achieve good volume reconstruction, thus promoting the dip transform to be a more practical technique.
The main problem with current SfS models, however, is that they artificially assume that a shape is full underneath its visible surface \cite{laurentini1994visual,kolev2009continuous}. Since SfS cannot ``see'' areas on the shape that are hidden from direct view, this assumption can potentially prevent us from obtaining an accurate assessment of an object's volume, while such hidden parts may be recovered well by the dip transform. Thus, to complement the dip transform we modify the SfS model to generate its data based only on the object's surface. To this end, we suggest to use a Softmax function with SfS model, which ensures that the SfS tracks the rays and generates the silhouettes based on a projection of the object's surface only. This procedure is summarized is Section \ref{sec:new_SfS}.

For the joint reconstruction, we combine the data misfit functions of the two modalities. Refering again to the example of Eq. \eqref{jointMisfitU}, we obtain:
\begin{equation}\label{jointMisfit}
\argmin_{\bfm_\joint} \left\{\sum_{j=1}^{n_\dip}\mbox{loss}^{\dip}_j(Q_\dip\bfm_\joint) + \gamma \sum_{l=1}^{n_\sfs} \mbox{loss}^{\sfs}_l(Q_\sfs\bfm_\joint) + \lambda R(\bfm_\joint)\right\},
\end{equation}
where $\bfm_\joint$ is the joint vector of parameters for the two problems, and therefore, it contains the set of PaLS parameters and separate sets of dip and SfS data acquisition parameters:
\begin{equation}
    \bfm_{\joint} =  \left[ {(\alpha_i,B_i, \xi_i)}_{i=1}^{n_{RBF}}, {(\theta_j , \phi_j , \vec{b}_j )}_{j=1}^{n_{\dip}},{(\theta_l , \phi_l , \vec{b}_l )}_{l=1}^{n_{\sfs}} \right].
\end{equation}
The matrices $Q_\dip$ and $Q_\sfs$ extract the relevant parameters from $\bfm_\joint$ for each of their corresponding modalities, and $\gamma> 0$ is used to balance the impact of the different objectives. Note that by using this approach we could, theoretically, combine more than two input channels in our problem. This joint reconstruction is easily set up by using the jInv inversion package \cite{jInv17}, which we use for our computations. Joint multi modal inversions using this package were also recently applied in a geophysical context \cite{JointEikFWI17,fung2018uncertainty}.

\subsection{No-fill Shape from Silhouettes}\label{sec:new_SfS}
The second input source is a modified version of the SfS algorithm that was suggested in \cite{kolev2009continuous}.
Each pixel in a picture taken by the camera is computed by a ray extending from the camera to that pixel, possibly hitting the object. We denote the constructing ray of the $(i,j)$-th pixel by $\bfr_{i,j}$, and its value by $d_{ij}$ ($\bfr_{ij}$ is the set of indices that constitute the ray).
Ideally, we would have a binary 3D object and a 2D silhouette, and if the ray $\bfr_{ij}$ hits the object's edge or surface, the model should output 1. That is, if we observe two adjacent voxels in the direction of the ray, wherein the value of the first is 0 and that of the other is 1, it indicates that we hit the surface of the object and should output 1. For a real-life silhouette simulation, the part of the object that lies beyond the surface does not influence the result produced by the model.

In practice, we need to work with a smooth transition from 0 to 1, and hence, any voxel that is near the boundary of the object registers a gradual change as it moves from the background to the object itself. Similarly to the ideal case, we wish to recognize this boundary layer and to output the maximum value in that layer for the silhouette. Additionally, we would like the silhouette value to depend on all the voxels in the boundary layer, an element that is of critical importance to the optimization process in the PaLS framework. To achieve such dependence, we define a voting process according to the ray's first increasing sequence of values $\bfu$, which indicates the rays encountered the object's boundary.

Our voting process for computing $d_{ij}$ entails performing a softmax of the values of $\bfu$ over the first boundary layer in the ray $\bfr_{ij}$ (increasing values of $\bfu$). That is:
\begin{equation} \label{eq:silhouette_simulation}
    d_{ij}(\bfu) = \sum_{k\in \Omega(\bfr_{ij})}{\frac{\bfu[\bfr_{ij}[k]] \exp{(\eta \bfu[\bfr_{ij}[k]]})}{\sum_{t\in \Omega(\bfr_{ij})}{\exp{(\eta \bfu[\bfr_{ij}[t]]})}  }},
\end{equation}
where $\Omega(\bfr_{ij})$ is the subset of indices that correspond to \emph{increasing} values in $\bfu$ in the ray $\bfr_{ij}$. The behavior of the softmax function is controlled by setting $\eta > 0$ ($\eta=50$ in our tests). We note that insofar, since the process in \eqref{eq:silhouette_simulation} is designed to be realistic and is defined only by the shape's boundaries, the model cannot determine the invisible inner parts of the shape. These parts are left to regularization only. Alternatively, as in our case, one can complement this model with a tomographic model like the dip transform to fill the missing inner information.

\subsection{Mesh-free implementation for reconstruction from point clouds} Suppose we are given a set of points $\mathcal{C} = \{\vec x_i\}_{i=1}^{N}$ that lie on the surface of the object. Because our PaLS framework provides a smooth transition from 0 to 1 using the Heaviside function in Fig. \ref{fig:heavyside}, our objective is to find the PaLS parameters $\bfm$, s.t. 
\begin{equation}\label{eq:levelset_PC}
\forall \vec{x}_i\in\mathcal{C}:\quad u(\vec x_i,\bfm) \approx \delta,
\end{equation}
where $\delta$ is our binarization threshold parameter in Alg. \ref{alg:Reconstruction}. In order to consistently define the inner and outer sides of the shape for all the points, we use the common technique of \cite{carr2001reconstruction}, as follows. Using the points in $\mathcal{C}$ we estimate surface normals $\{\vec n_i\}_{i=1}^N$, and define two additional sets: $\mathcal{C}_\fwd = \{\vec x_i+\epsilon \vec n_i\}_{i=1}^{N}$, and $\mathcal{C}_\bwd = \{\vec x_i-\epsilon \vec n_i\}_{i=1}^{N}$. Then, in our loss function we try to fit (in a least squares sense) $u(\vec x_i,\bfm)=0$ for $\vec x_i\in\mathcal{C}_\fwd$, and $u(\vec x_i,\bfm)=1$ for $\vec x_i\in\mathcal{C}_\bwd$, in addition to \eqref{eq:levelset_PC}. 

Using the approach above, our loss depends only on the points defined in the three given sets, and there is no need to define a volumetric grid as in the other problems. Hence, we use our PaLS model with a collection of points, in a mesh-free manner. That is, the parameters $\bfm$ are estimated using a mesh free minimization of the loss over the points in the clouds, and the shape can be constructed later on an arbitrarily fine mesh. The value of $\epsilon$ is chosen according to the width of the transition from 0 to 1 in the Heaviside function. In addition, to have a fast implementation, the procedure that computes $u(\bfm)$ has to efficiently determine which points in the clouds are relevant for each basis function. We obtain this using a nearest neighbor data structure implemented in the Julia package {\tt NearestNeighbors.jl}, which is available on GitHub.     

\section{Results}\label{sec:results}
To validate our proposed framework for 3D shape reconstruction, we conducted several experiments. The first experiment aims to demonstrate that even though the PaLS framework contains a representation using smooth RBFs, it can also be used to represent sharp objects, which are typically difficult to define using basis functions. In the subsequent experiments, we exploit the dip transform model for tomographic data acquisition and compare the reconstructions obtained with our framework to those obtained with the original reconstruction method in \cite{aberman2017dip}. In our third batch of experiments, we demonstrate the robustness of the reconstruction algorithm to uncertainty in the acquisition parameters. Following that, we show that multi-modal shape reconstruction is possible with our framework via the combination of the modified SfS model described in Section \ref{sec:new_SfS} with the dip transform model. This experiment not only shows that we are able to perform a joint shape reconstruction, it also demonstrates how the large number of experiments required for the dip transform can be substantially reduced. Finally, we show the reconstruction of an object from two given misaligned and non-overlapping point-clouds, and show that their combination with SFS can lead to a correct reconstruction.

All the shape reconstructions are obtained using Gauss-Newton in Algorithm \ref{alg:Reconstruction} with iterated Tikhonov and log-determinant barrier regularization. Our framework is implemented in the Julia language \cite{Julia}, and we use the inversion package jInv \cite{jInv17} to perform the computations, which are parallelelized over the experiments.
We use MeshLab as a final step to post-process the volumetric meshes we obtained. The post-processing steps include Laplacian-smoothing and voxel sub-sampling to avoid artifacts, which can be caused by over-sampling the continuous representation obtained using our framework. Our code for reproducing the results in this work is available online at:
\begin{center}{\tt
\url{https://github.com/BGUCompSci/ShapeReconstructionPaLS.jl}}
\end{center}

\subsection{Shape Representation using PaLS}
\begin{figure}
    \centering
     \begin{subfigure}[b]{0.22\textwidth}
        \includegraphics[width=\textwidth,keepaspectratio,trim={1cm 5.0cm 4cm 3cm},clip]{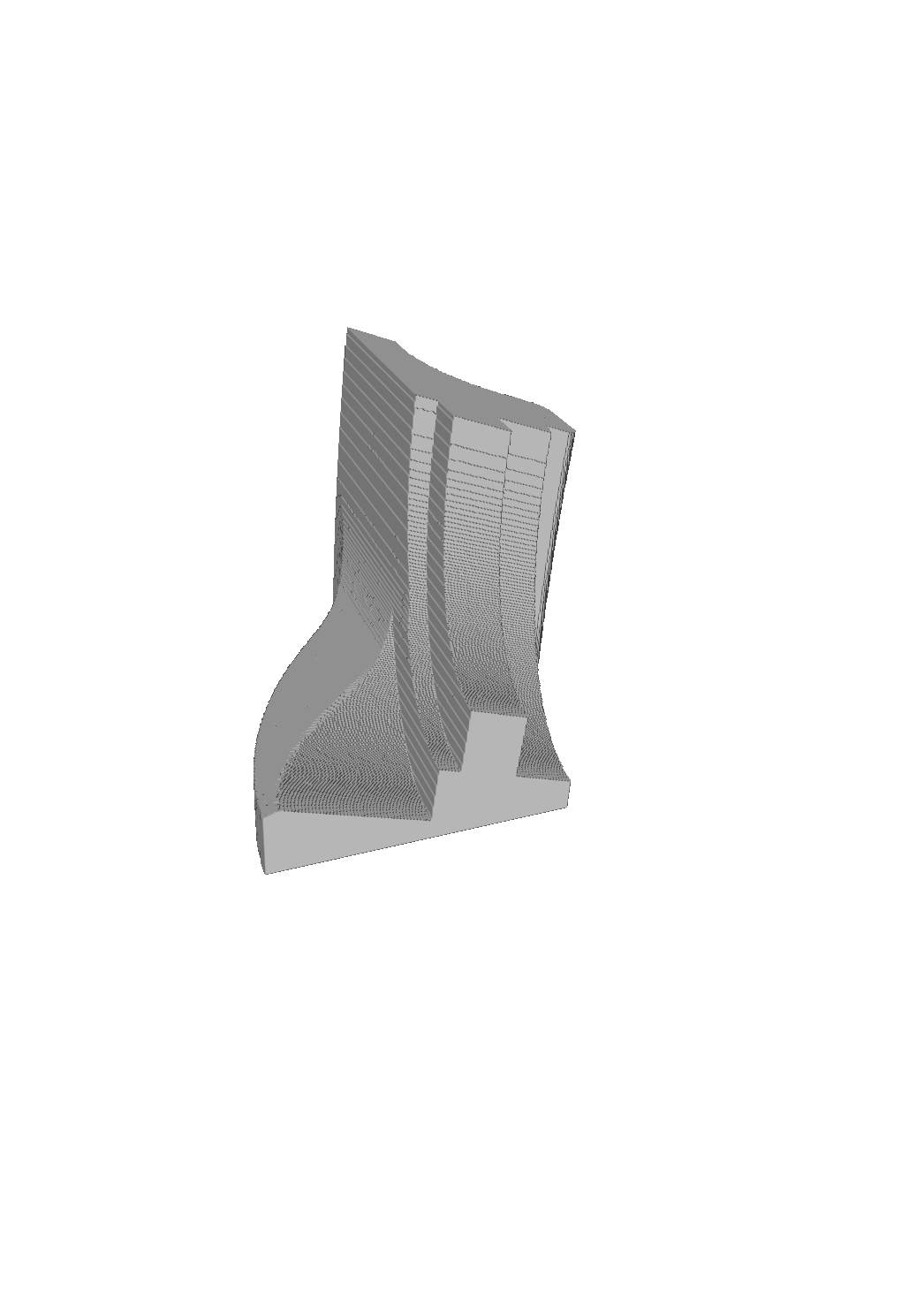}
        \caption{Input object}\label{fig:fandisk}
    \end{subfigure}%
    \begin{subfigure}[b]{0.22\textwidth}
        \includegraphics[width=\textwidth,keepaspectratio,trim={1cm 5cm 4cm 3cm},clip]{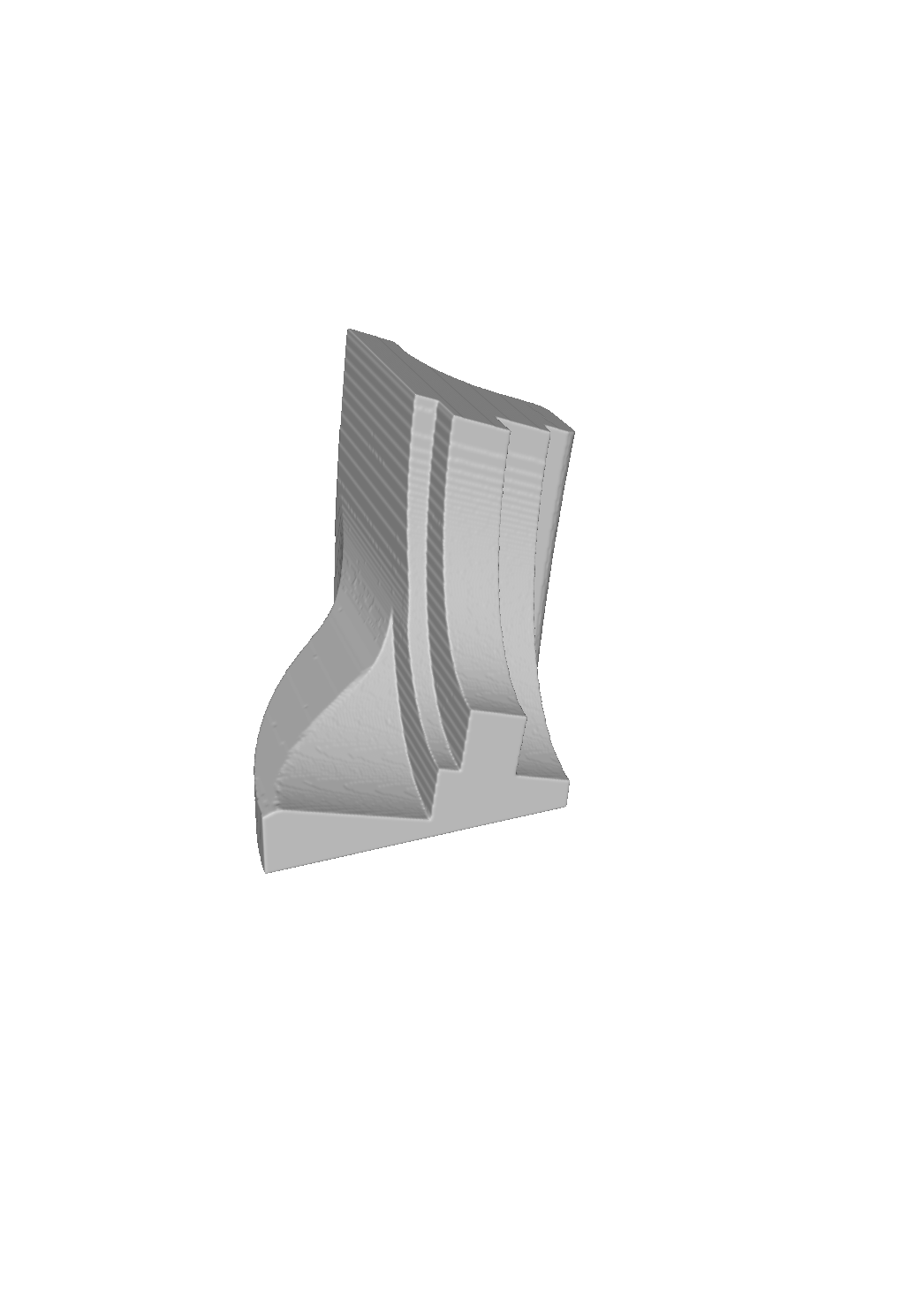}
        \caption{Smoothed Input}
    \end{subfigure}%
    \begin{subfigure}[b]{0.19\textwidth}
        \includegraphics[width=\textwidth,keepaspectratio,trim={18cm 8.1cm 28cm 8cm},clip,angle=-3,origin=c]{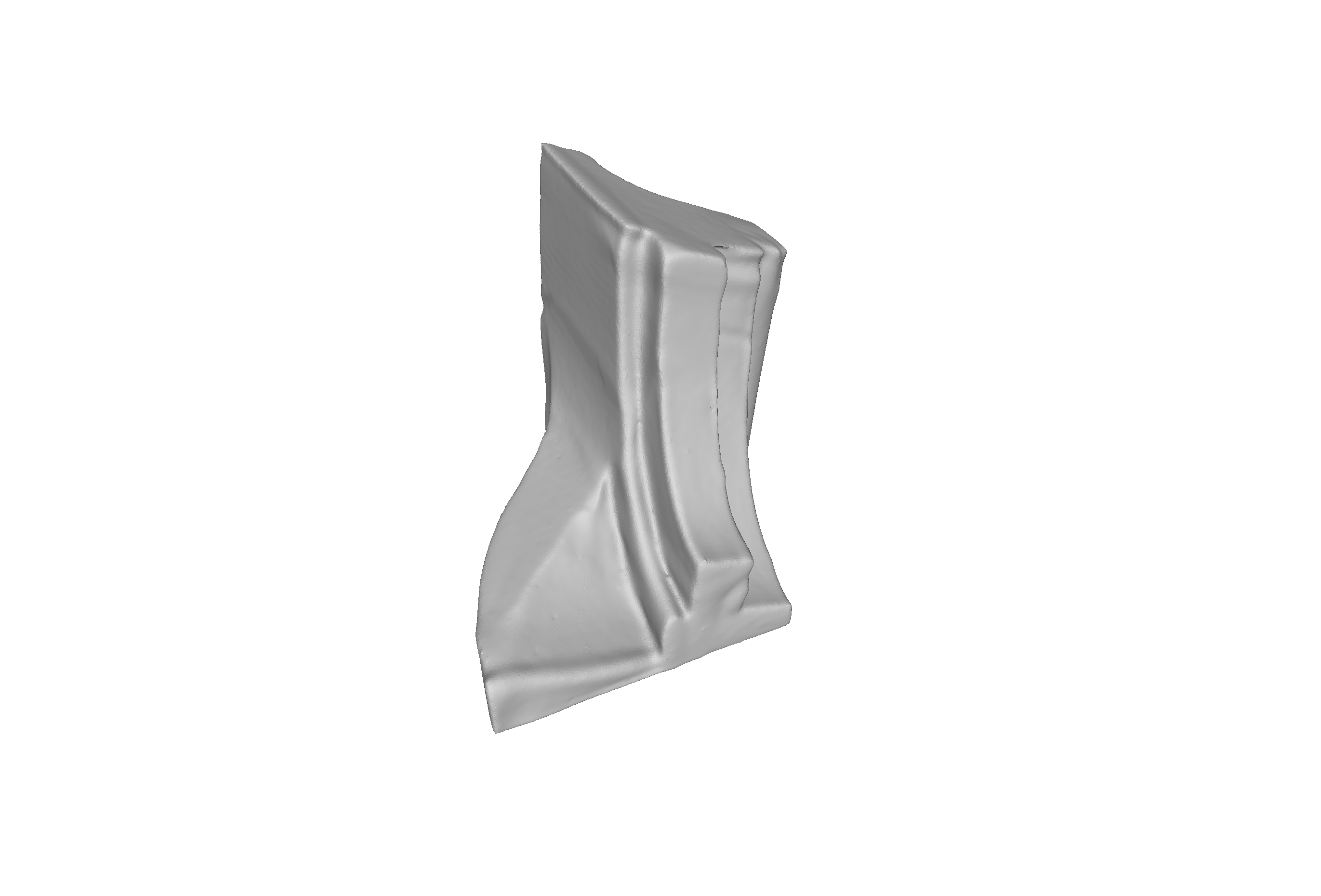}
        \caption{78 RBFs}
    \end{subfigure}%
     \begin{subfigure}[b]{0.19\textwidth}
        \includegraphics[width=\textwidth,keepaspectratio,trim={1cm 4.5cm 4.0cm 3cm},clip]{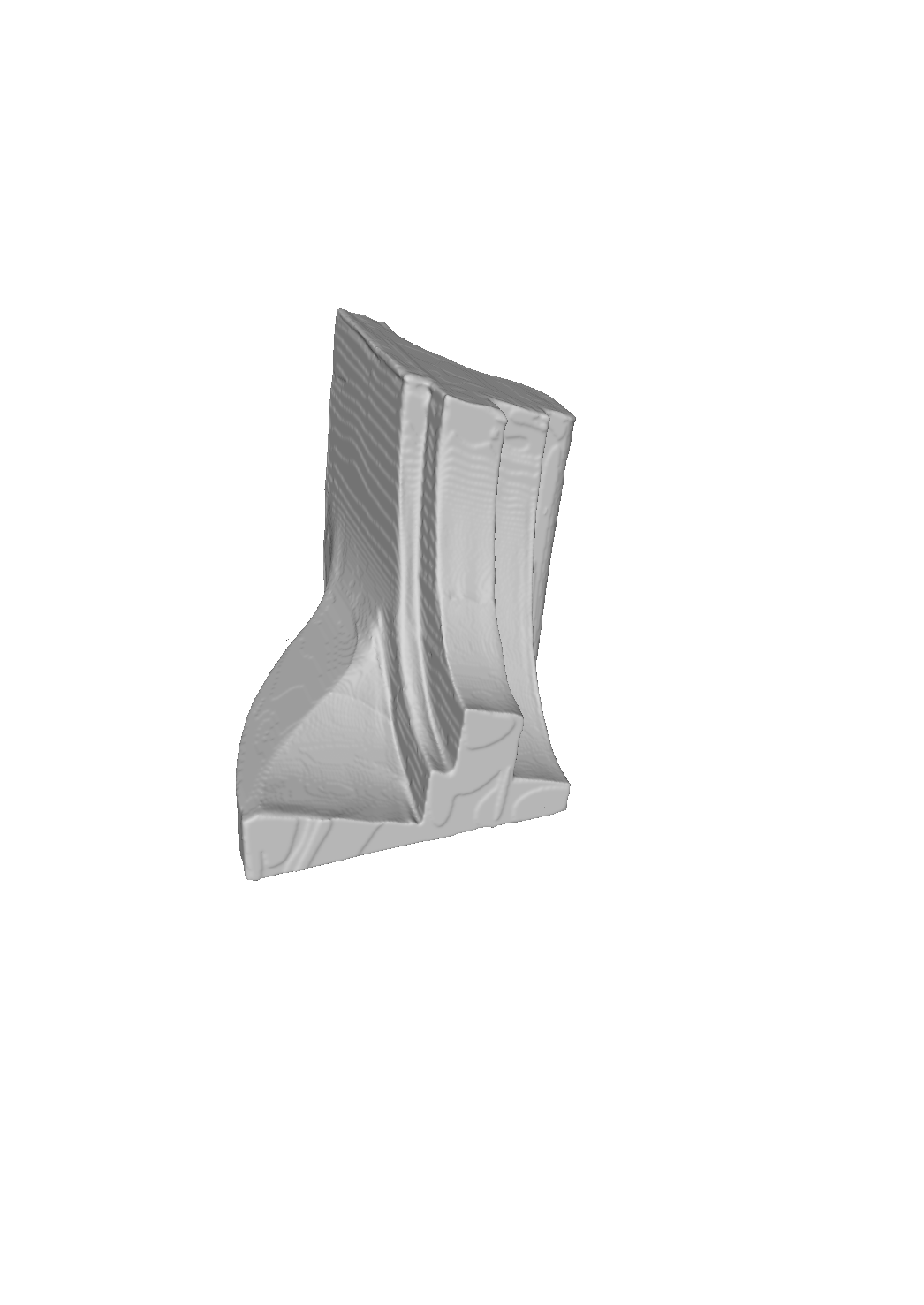}
        \caption{120 RBFs}
    \end{subfigure}
\caption{Sharp object representation. (a) The volumetric input as it is consumed by our framework. (b) Smoothed version of the volumetric input. (c) and (d) Reconstructions using 78 and 120 RBFs, respectively.}
\label{fig:directResults}
\end{figure}

The outcome of this experiment serves as a proof of concept for the modeling capabilities of the ellipsoidal PaLS. Given an object $\bfu$ defined on a mesh (of size $120\times120\times120$), we search for $\bfu(\bfm)$ and assess the quality with which the PaLS representation models the object (based on a moderate number of RBFs). Since our goal is to enable objects to be modeled with the lowest possible number of parameters, we initialize the reconstruction with 5 RBFs and add only one RBF at each iteration. The results of this experiment can be seen in Fig. \ref{fig:directResults}.
We intentionally chose an input object with both flat areas and sharp angles to demonstrate the expressiveness of our model. The results show that our model can faithfully reconstruct both the sharp and the flat elements, tasks that are considered particularly challenging in shape reconstruction. Furthermore, despite the relative complexity of the shape, its reconstruction is done using a small number of parameters. As expected and can be seen in Fig. \ref{fig:directResults}, increasing the number of basis functions (while maintaining a relatively low number of parameters) improves the reconstruction.

\bigskip

\paragraph{Shape Reconstruction Setting and Default Parameters}
In Sections \ref{sec:CompWithOrig}-\ref{sec:dipVisResults} we present a few shape reconstruction experiments. To make the experiments realistic, all the data measurements are generated from a binary object using a $240\times240\times240$ grid. The data are then down-sampled, and we perform the reconstruction on a $120\times120\times120$ grid. We also include uncertainty in the acquisition parameters (detailed later) and Gaussian white noise of variance $\sigma = 2V$, where $V$ is the volume of a voxel.
For the reconstruction, we use algorithm \ref{alg:Reconstruction} and its default parameters. That is, in all the experiments, we start with 20 randomly located RBFs ($p_0 = 20$), and at each iteration, we add 5 basis functions to improve the reconstruction ($p=5$). We perform 40 iterations that result in 220 radial basis functions, and a typical reduction of 3 order of magnitude in the data term. The value of the regularization $\lambda$ is initially chosen to be $10^{-3}$ and we reduce it by a factor of $0.8$ after each iteration. Note that in every GN iteration, the regularization is zeroed according to the iterated Tikhonov approach described in Section \ref{sec:palsSection}. 
We set the grid domain to be the $[0,5]^3$ cube. Each of the initial RBFs is initialized as sphere with a radius of 1 (i.e., $B_i = I_{3 \times 3}$), located randomly around the center of the grid. Its coefficient $\alpha_i$ is initialized with a small random number. When adding RBFs during the optimization process, we set them with a smaller radius of $\frac{1}{3}$, and zero their coefficient $\alpha_i$, such that we obtain monotonically non-increasing optimization routine. The location of the added RBFs is determined via the gradient of the objective function w.r.t $\bfu$, such that new basis functions are added at grid points where the magnitude of the gradient is largest to improve the representation of the object where it lacks the most.

\subsection{Volumetric vs. PaLS Reconstruction} \label{sec:CompWithOrig}

\begin{figure}
    \centering
\begin{tabular}{|c|c|c|c|}
  \hline
  \multicolumn{2}{|c}{Known calibration} & \multicolumn{2}{|c|}{Uncertain calibration}\\
  \hline
  PaLS  & TV & PaLS & TV \\
  \hline
   \includegraphics[width=0.22\textwidth,keepaspectratio,trim={15cm 3.5cm 15cm 4cm},clip]{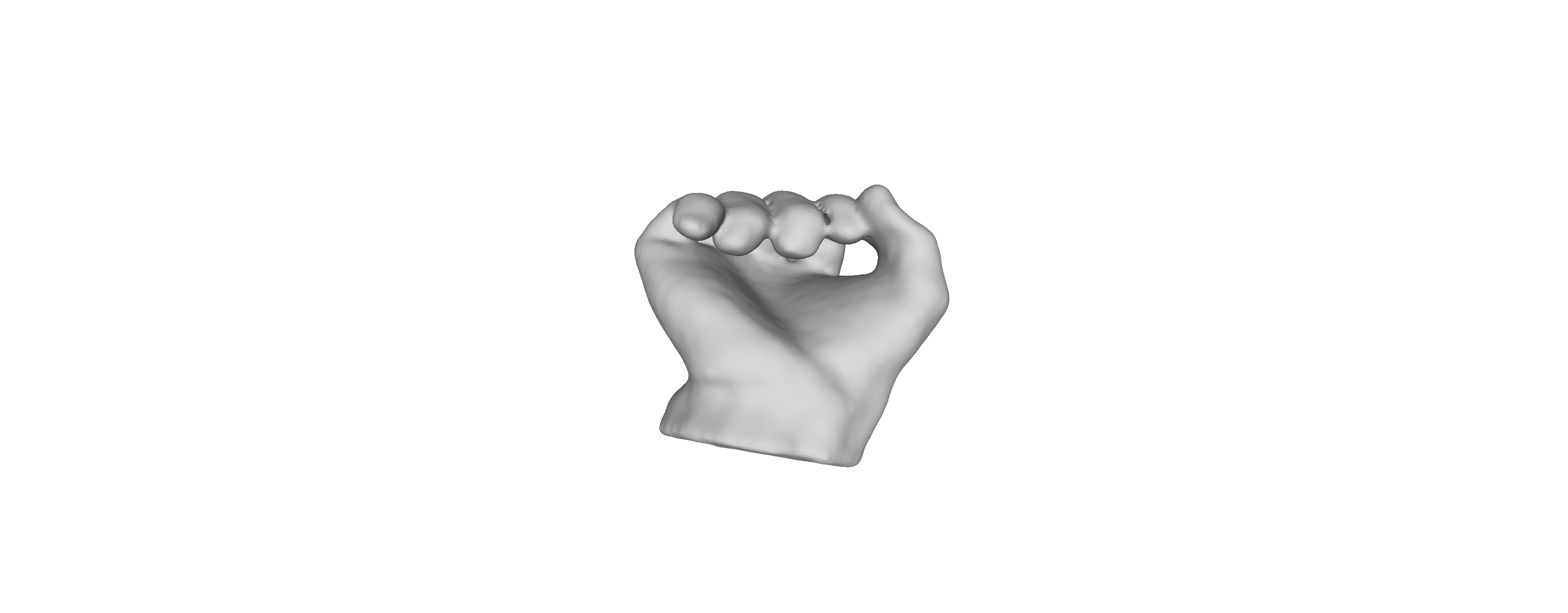} &
   \includegraphics[width=0.22\textwidth,keepaspectratio,trim={15cm 3.5cm 14cm 4cm},clip]{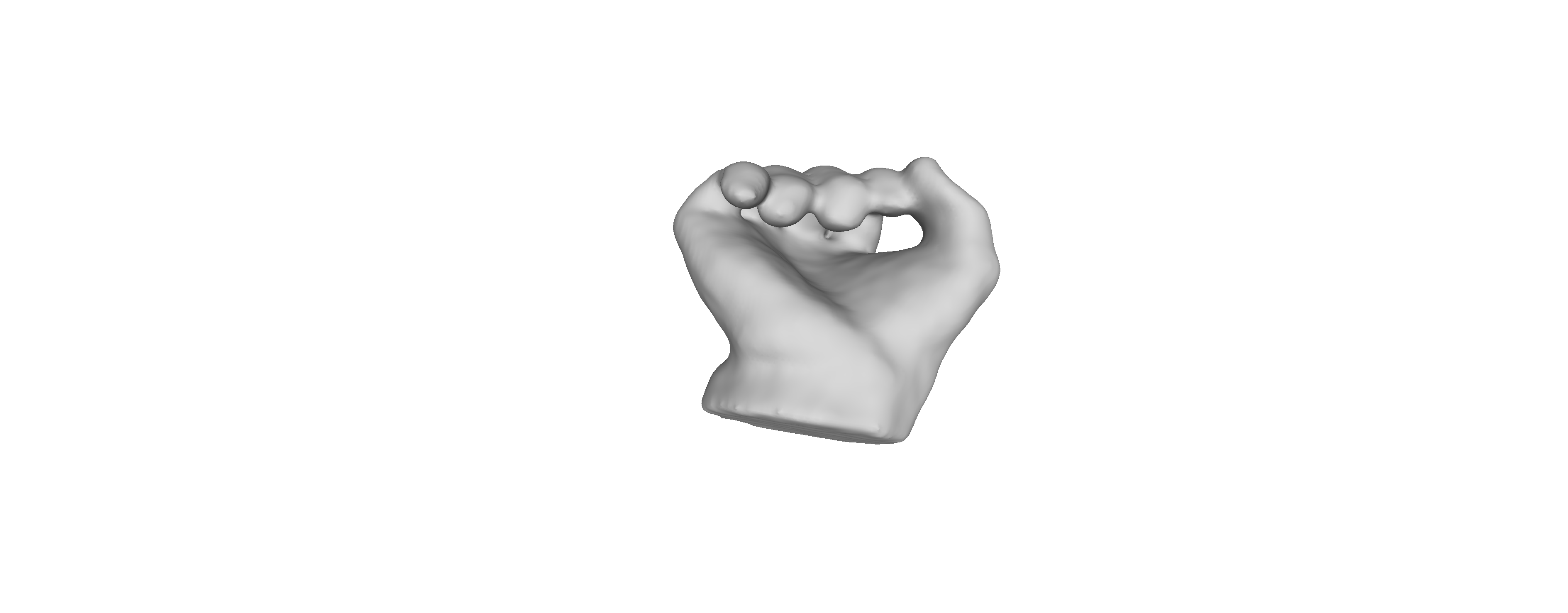}&
   \includegraphics[width=0.22\textwidth,keepaspectratio,trim={15cm 3.5cm 15cm 4cm},clip]{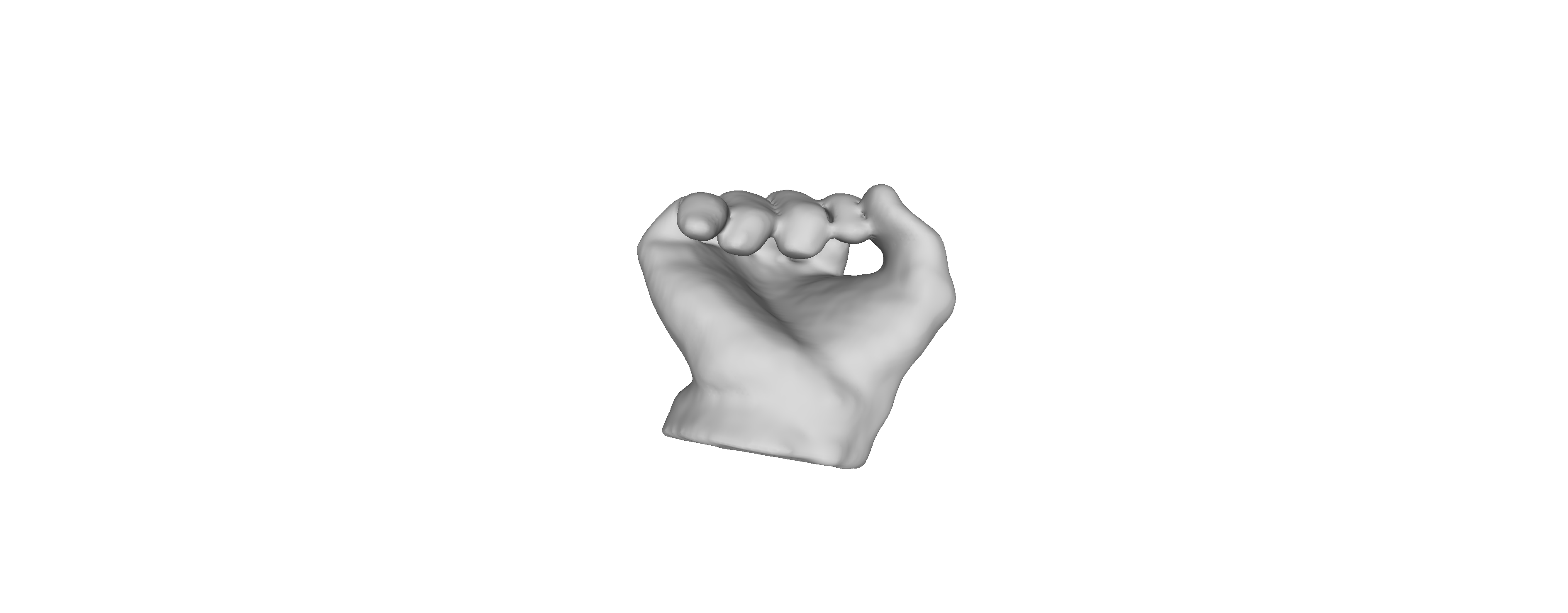}&
   \includegraphics[width=0.22\textwidth,keepaspectratio,trim={16cm 4cm 14cm 3.5cm},clip]{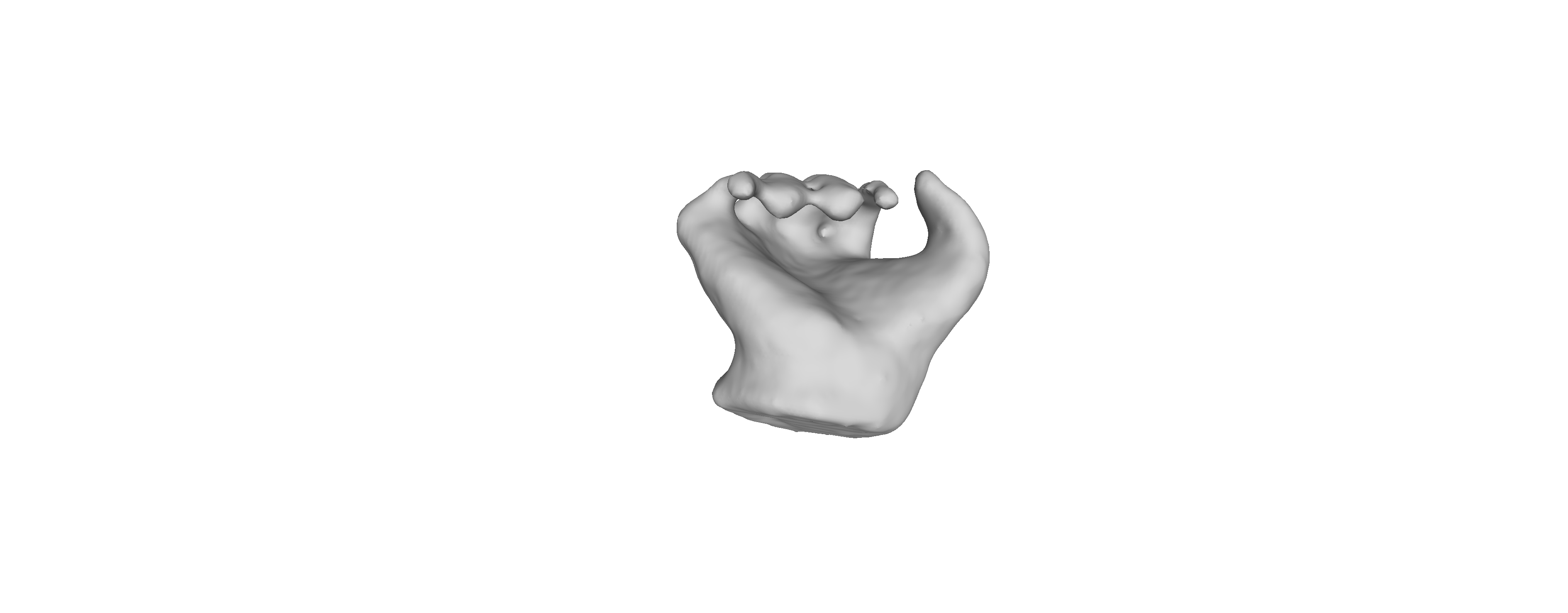}\\\hline
   \includegraphics[width=0.22\textwidth,keepaspectratio,trim={15cm 3.5cm 15cm 4cm},clip]{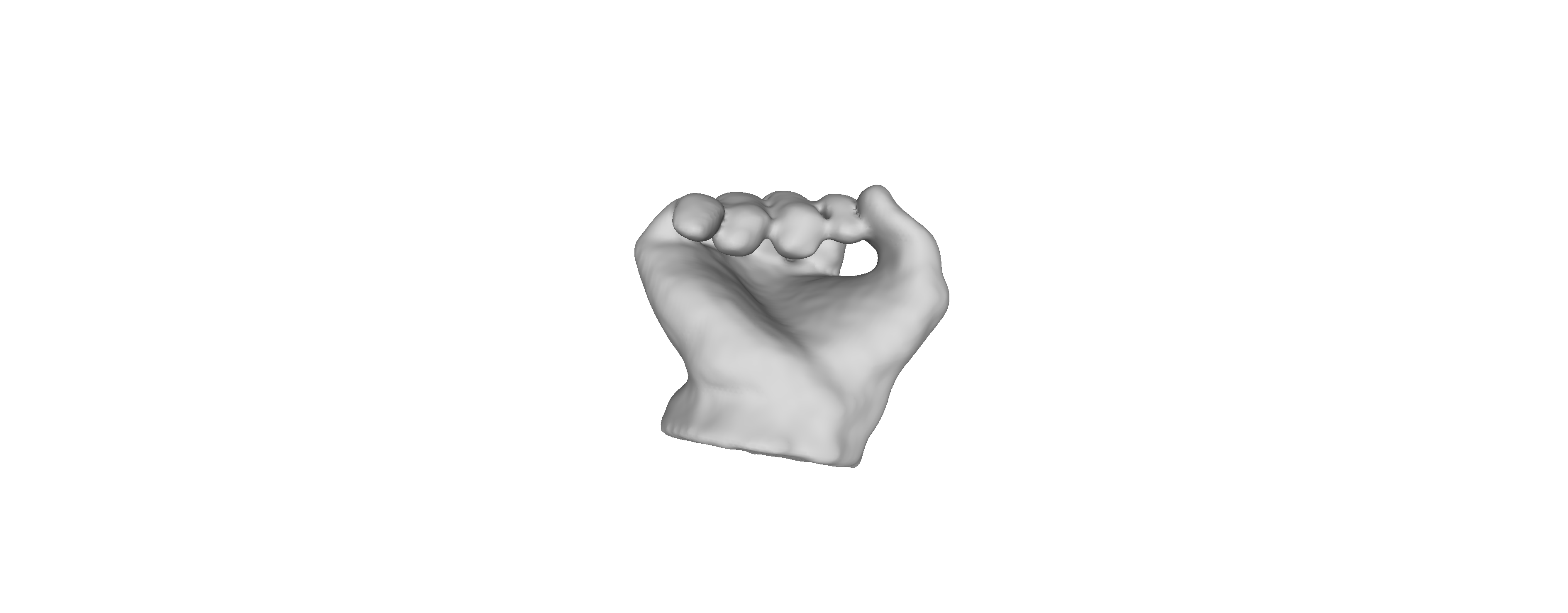}&
   \includegraphics[width=0.22\textwidth,keepaspectratio,trim={15cm 3.5cm 15cm 4cm},clip]{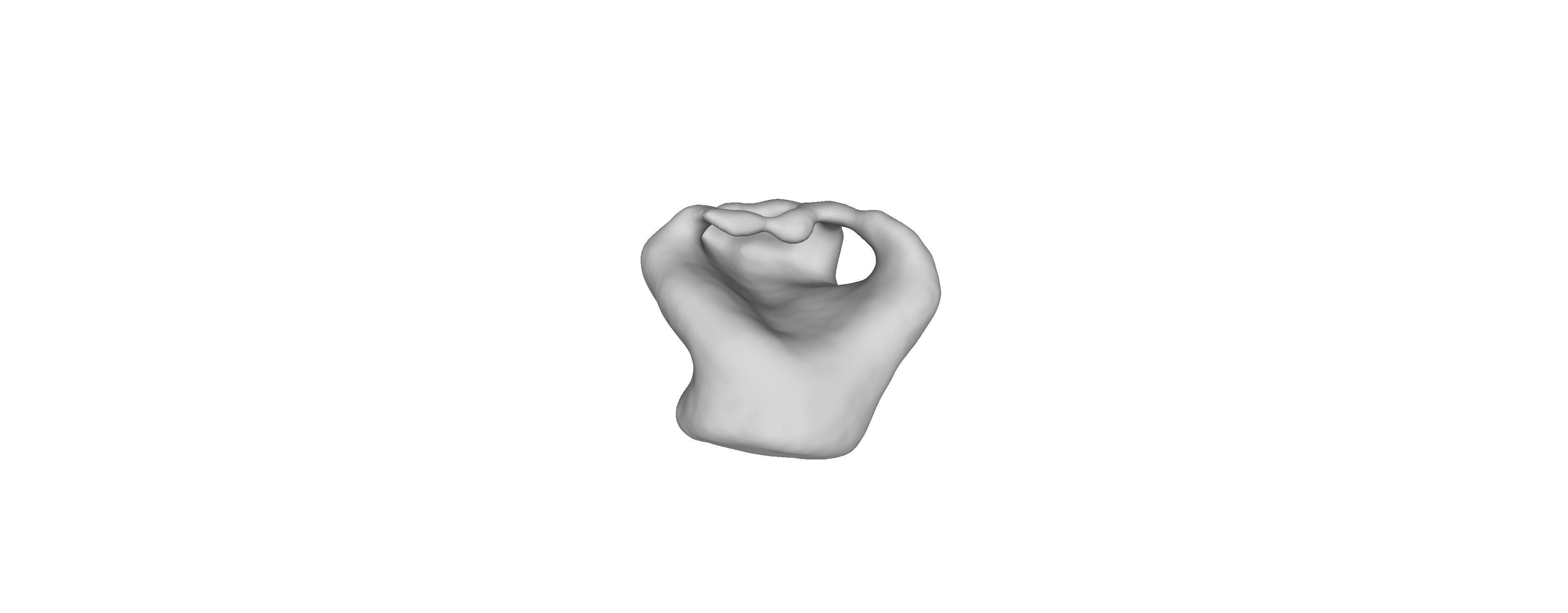} & 
  \includegraphics[width=0.22\textwidth,keepaspectratio,trim={15cm 3.5cm 15cm 4cm},clip]{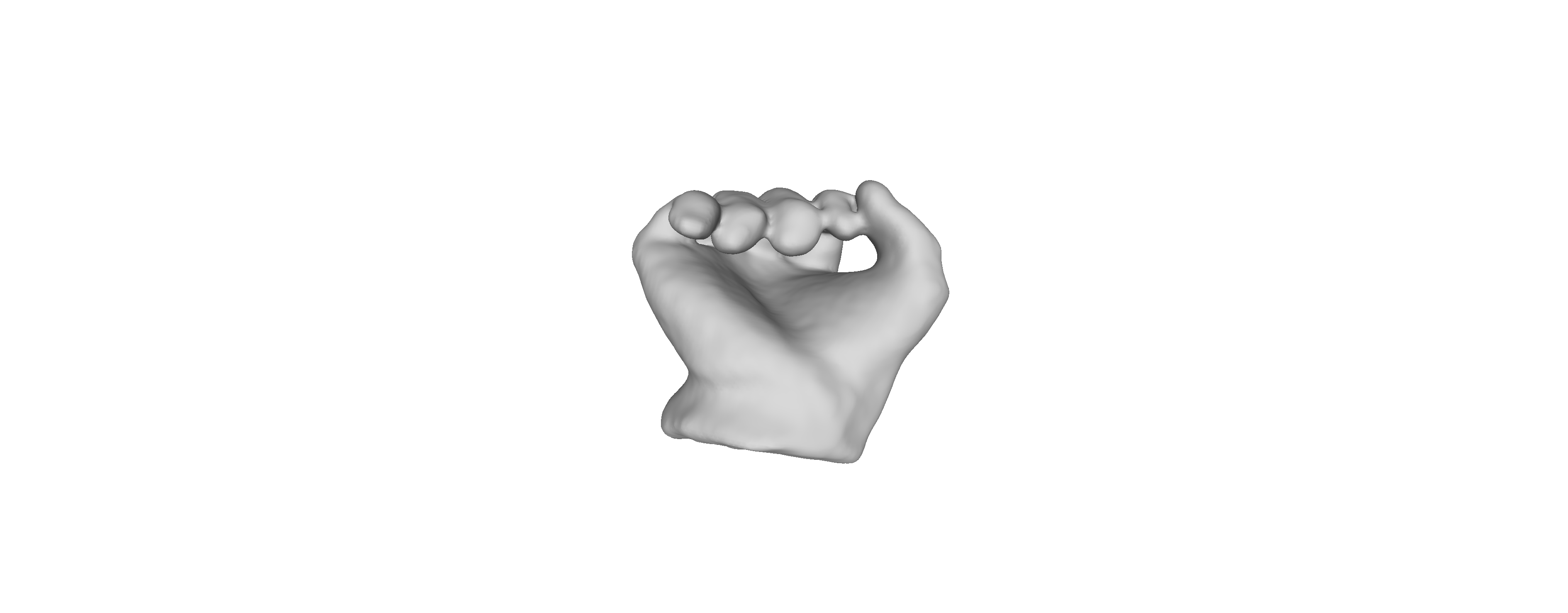}& \includegraphics[width=0.22\textwidth,keepaspectratio,trim={16cm 4cm 14cm 3.5cm},clip]{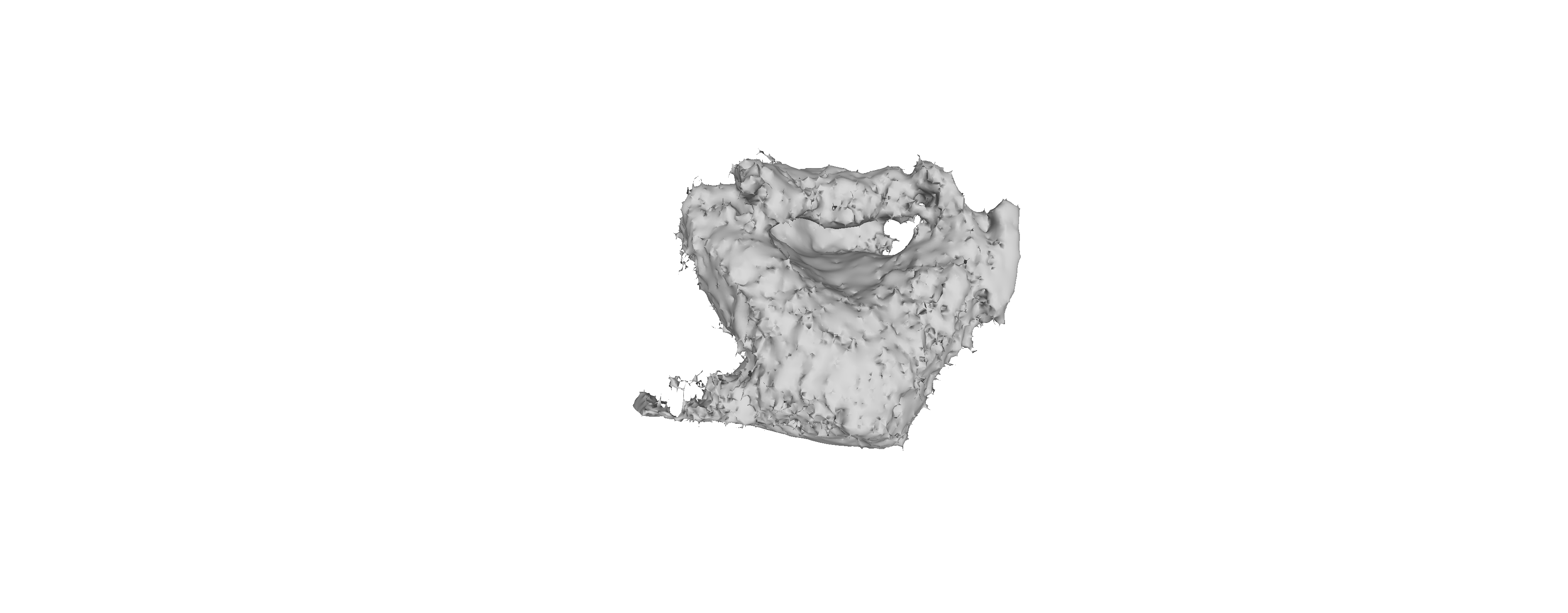}\\\hline
\end{tabular}
 \caption{Comparison between dip transform reconstruction similar to \cite{aberman2017dip} and our method. Noisy data consists of 2 degrees and 4\% translation noise. Top: reconstruction from 1000 dips (large data). Bottom: reconstruction from 400 dips (small data). All the experiments also include i.i.d. Gaussian white noise.}
    \label{fig:TV_vsRBF}
\end{figure}

In this batch of experiments, we demonstrate that our framework can significantly enhance object reconstruction when there is not enough data. We compare the original work  \cite{aberman2017dip} using TV regularization and our RBF-based dip transforms for different numbers of dips (small vs. large data) and different degrees of noise in the acquisition parameters. The results are summarized in Fig. \ref{fig:TV_vsRBF}. First, to accurately reconstruct the object, the number of dips required in our framework (400) is significantly lower compared to the original method, which requires at least 1000 data measurements to produce a good reconstruction (this is aligned with the results in \cite{aberman2017dip}). Secondly, and perhaps more importantly, our method is able to overcome uncertainty in the acquisition parameters, which, in the case of small data, completely ruins the TV reconstruction. In contrast to the tremendous impact that the calibration uncertainty has on reconstruction when using the conventional method, when using our method, its effect is small to insignificant.

\subsection{Robustness to Calibration Uncertainty}
\label{sec:RobustNoise}

\begin{figure}
    \centering
\begin{tabular}{c|c|c}
   a & b & c \\ \hline
   \includegraphics[width=0.3\textwidth,height=0.3\textheight,scale=4,trim={17cm 3cm 17cm 3cm},clip,keepaspectratio]{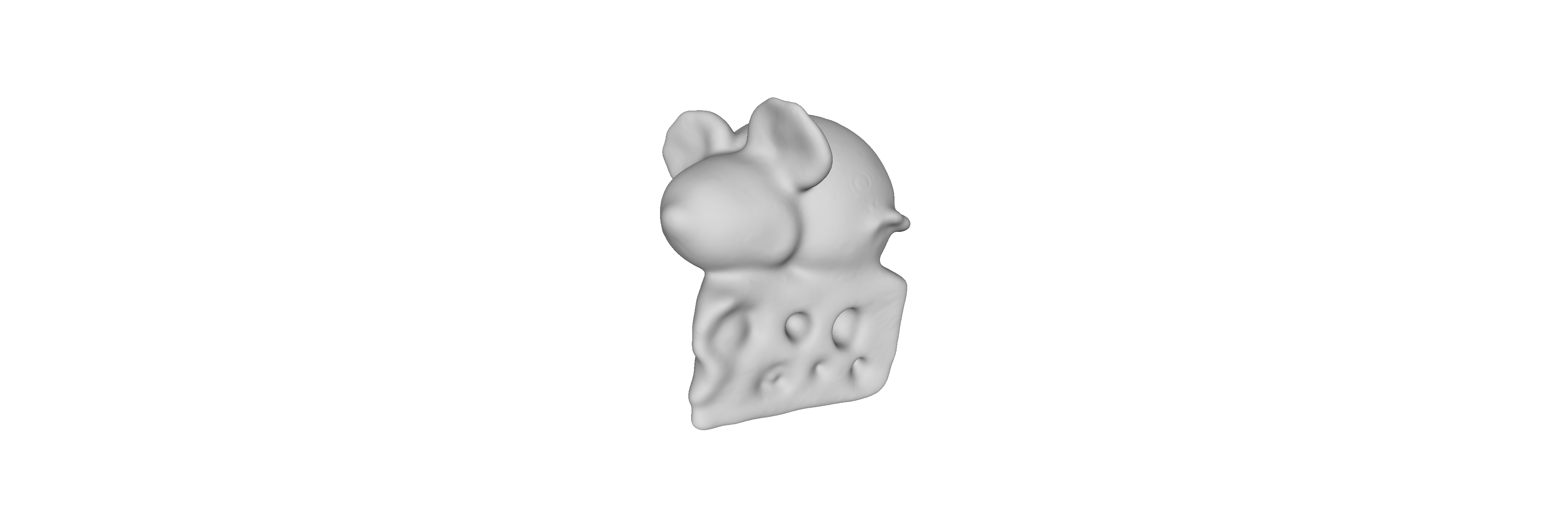} & \includegraphics[width=0.3\textwidth,height=0.3\textheight,scale=4,trim={16cm 2cm 16cm 2cm},clip,keepaspectratio]{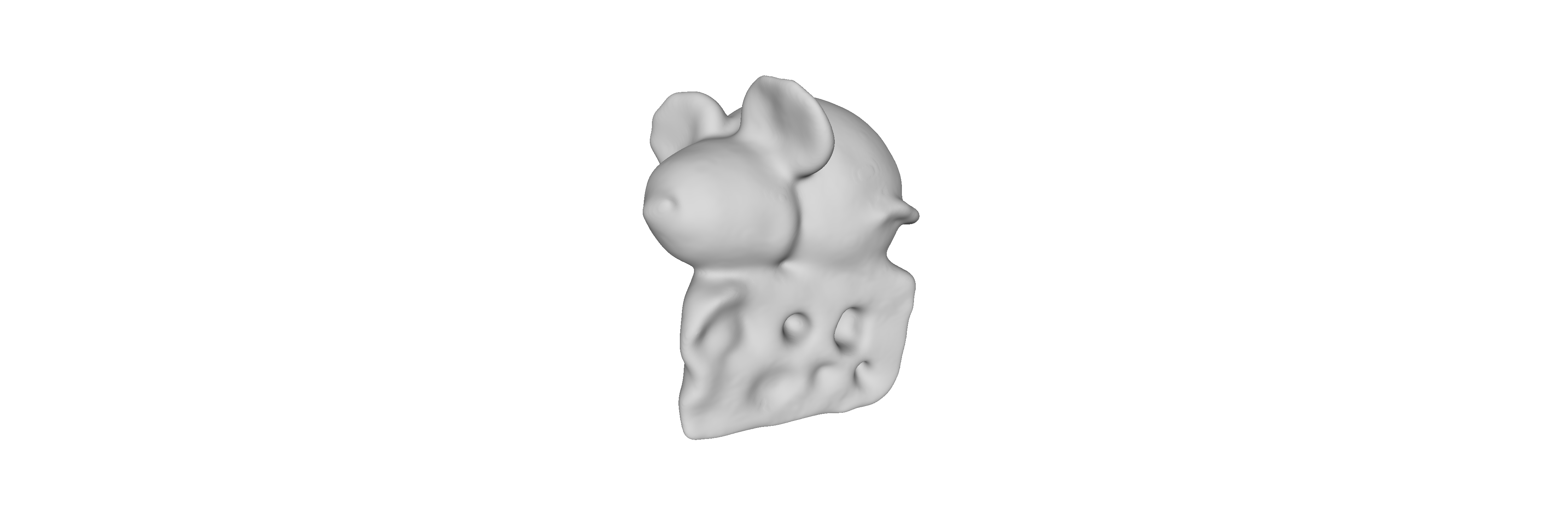} & \includegraphics[width=0.3\textwidth,height=0.3\textheight,scale=4,trim={16cm 2cm 16cm 2cm},clip,keepaspectratio]{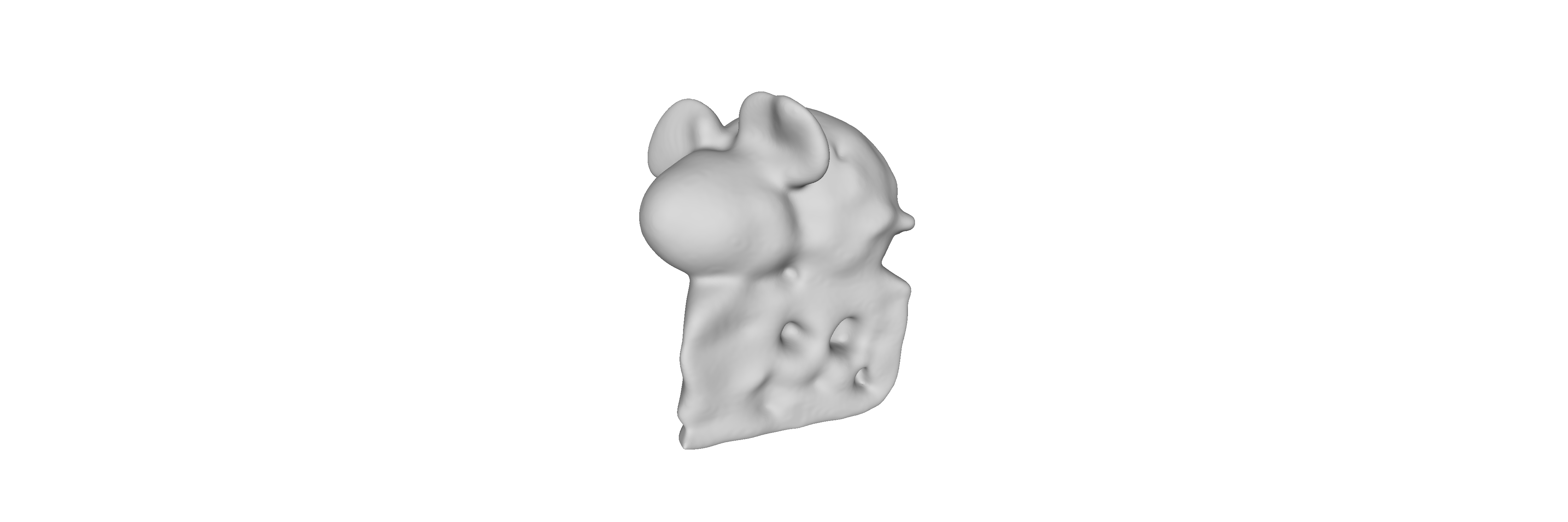} \\ \hline
    \includegraphics[width=0.3\textwidth,height=0.3\textheight,scale=4,trim={16cm 2cm 16cm 2cm},clip,keepaspectratio]{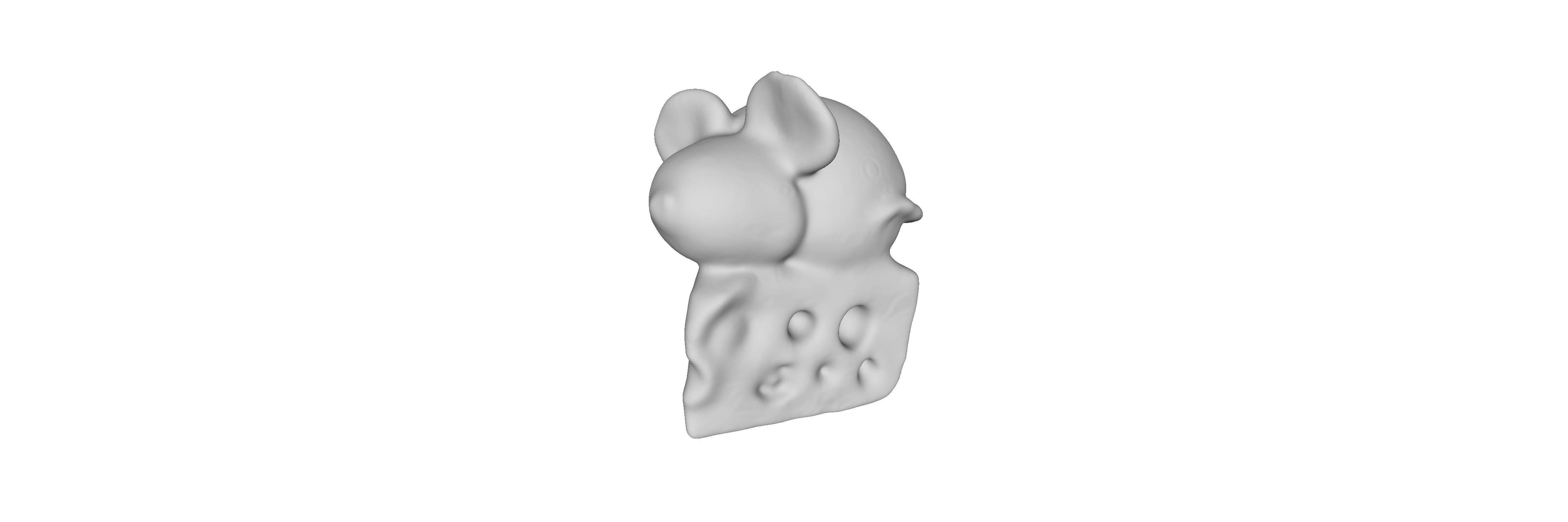} & \includegraphics[width=0.3\textwidth,height=0.3\textheight,scale=4,trim={16cm 2cm 16cm 2cm},clip,keepaspectratio]{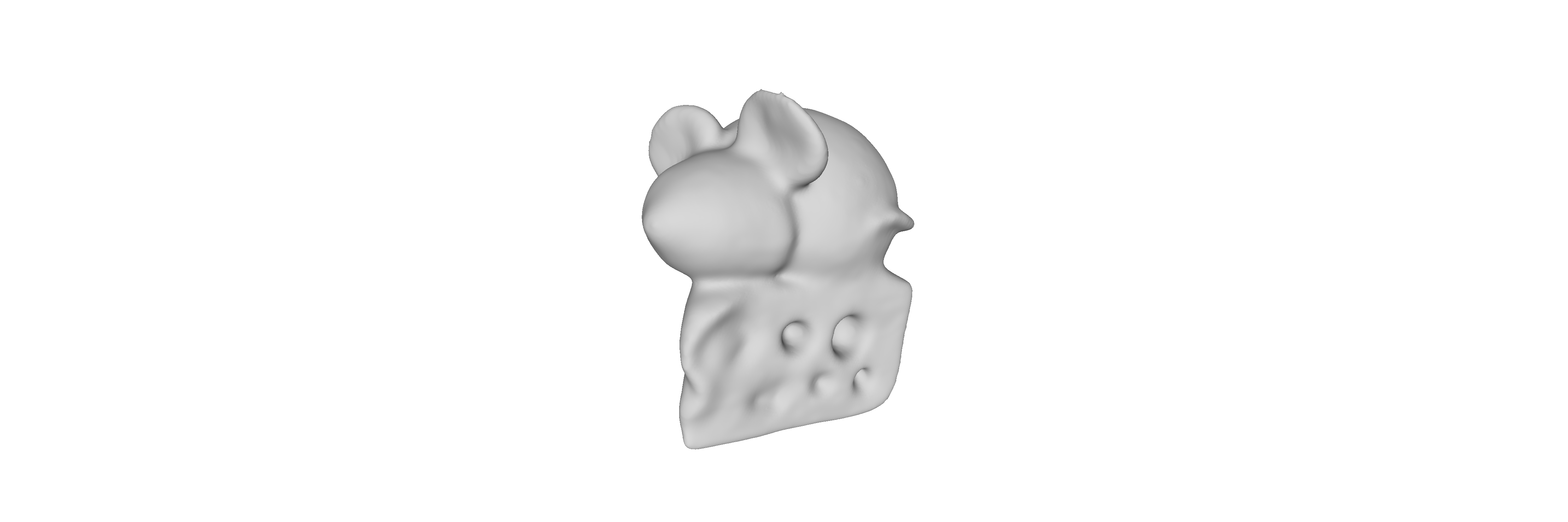} & \includegraphics[width=0.3\textwidth,height=0.3\textheight,scale=4,trim={16cm 2cm 16cm 2cm},clip,keepaspectratio]{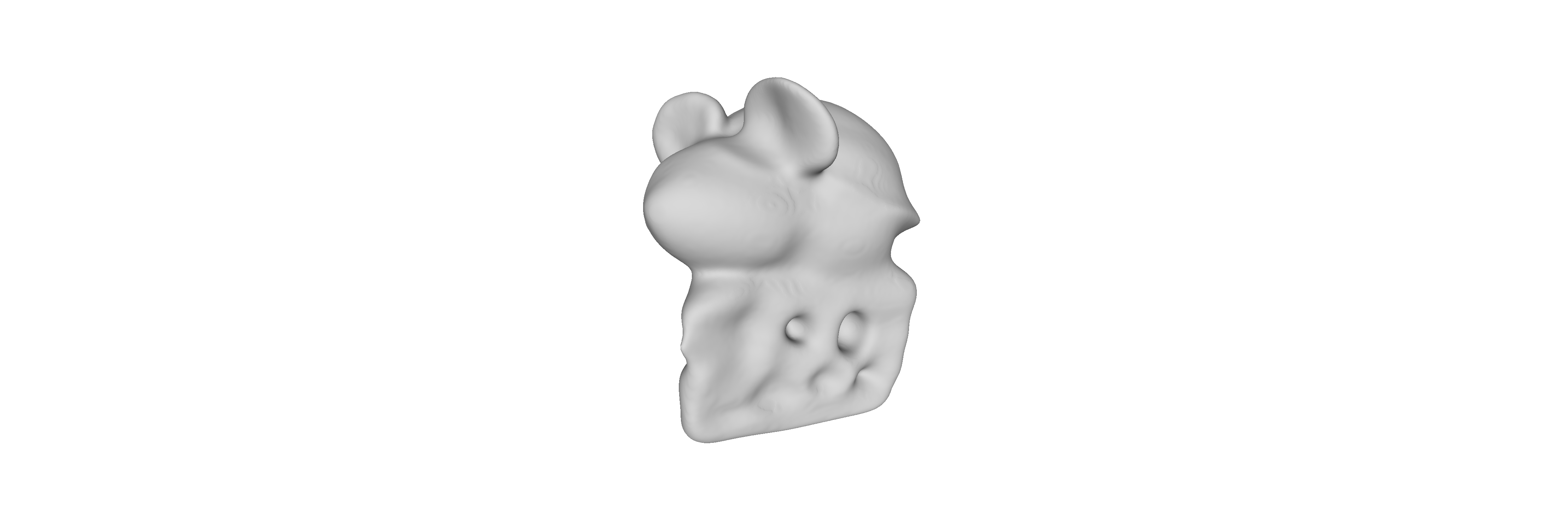}
\end{tabular}
 \caption{Robustness to calibration uncertainty in the PaLS reconstruction. First row: 200 dips. Second row: 600 dips. (a) Known calibration. (b) 4 degrees noise and 7\% translation noise (c) 10 degrees noise and 17\% translation noise. }
    \label{fig:noiseResults}
\end{figure}

In this section, we demonstrate the robustness of our reconstruction method to calibration uncertainty as discussed in Section \ref{sec:noiseHandling}.
We start with an experiment of a  reconstruction  assuming all the acquisition parameters are known. Then, we gradually increase the calibration uncertainty level until it degrades the results to the extent that the reconstruction no longer resembles the object.
Fig. \ref{fig:noiseResults} shows that our framework is capable of handling a significant amount of calibration noise---more than 10 degrees of noise in the object's rotation and 17 percent in its translation. To assess the effect of the data size on the final reconstruction and robustness to noise, we conduct the experiment for low (200) and high (600) numbers of measurements. As expected, as the number of measurements increases, the reconstructions become more robust to calibration noise.

\begin{figure}
    \centering
        \includegraphics[width=\textwidth,keepaspectratio]{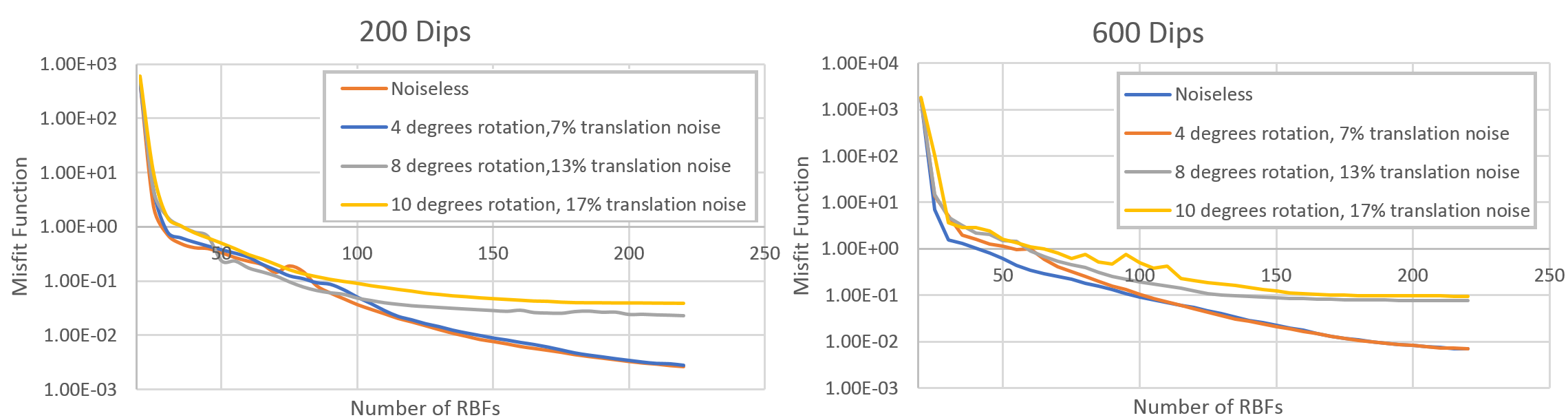}
\caption{Data misfit histories of the reconstruction with 200 and 600 dips (shown in Fig. \ref{fig:noiseResults}), with various noise levels in the acquisition parameters .}
\label{fig:noiseGraphs}
\end{figure}

Here we also show the decay of the loss function as we increase the number of RBFs to reconstruct the object (Fig. \ref{fig:noiseGraphs}). Note that the loss values do not include the regularization term, and hence they are not necessarily monotonically non-increasing. Our optimization method guarantees a monotonically non-increasing solution for the objective in Eq. \eqref{eq:invProblem_mrot}.

\subsection{Multi-modal Joint Reconstruction}\label{sec:dipVisResults}
\begin{figure}
    \centering
\begin{tabular}{c|c|c|c}
   a&b&c&d\\\hline
    \includegraphics[width=0.22\textwidth,keepaspectratio,trim={2.2cm 2.5cm 5cm 8cm},clip]{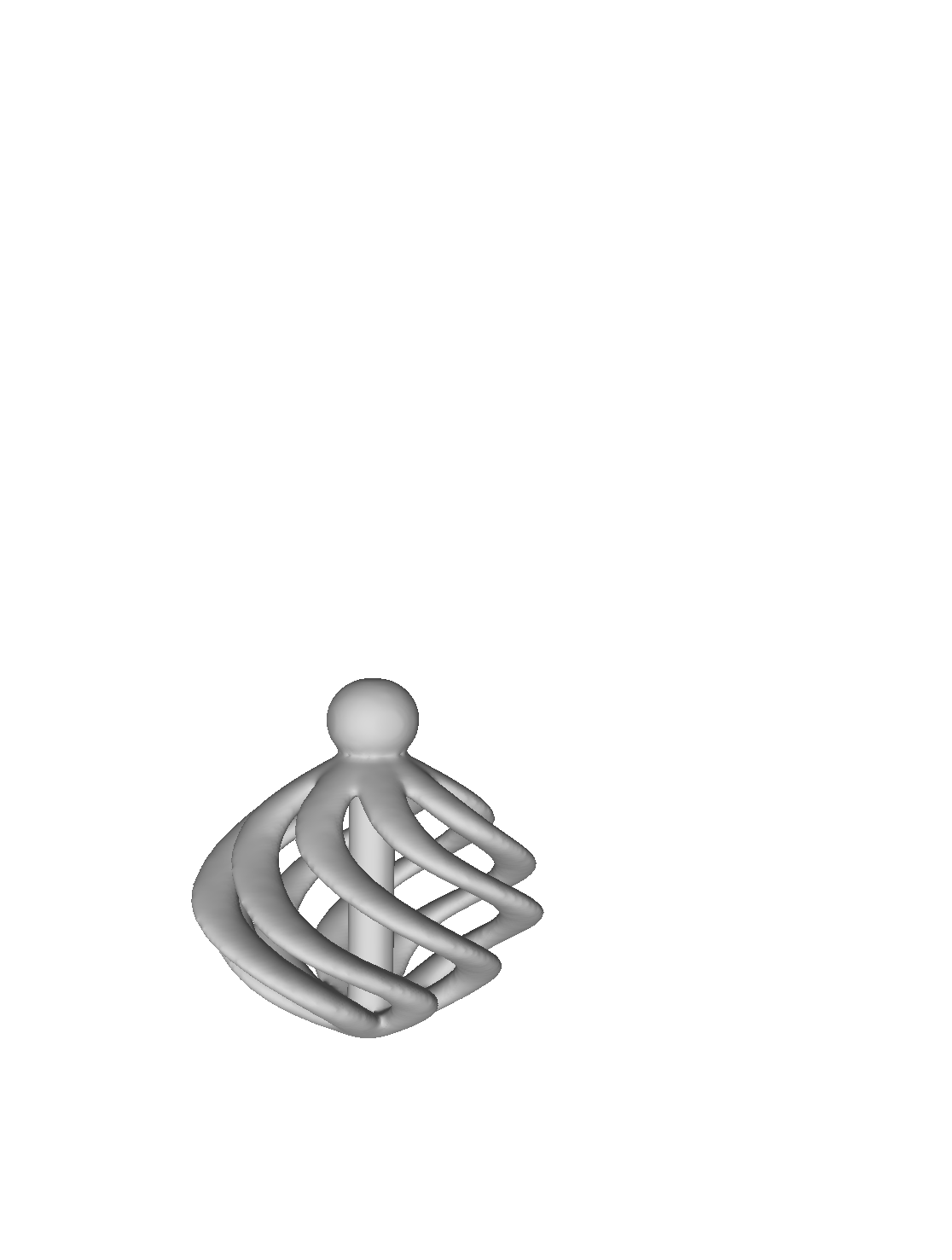}&
   \includegraphics[width=0.22\textwidth,clip,keepaspectratio,trim={2.2cm 2.5cm 5cm 8cm},clip]{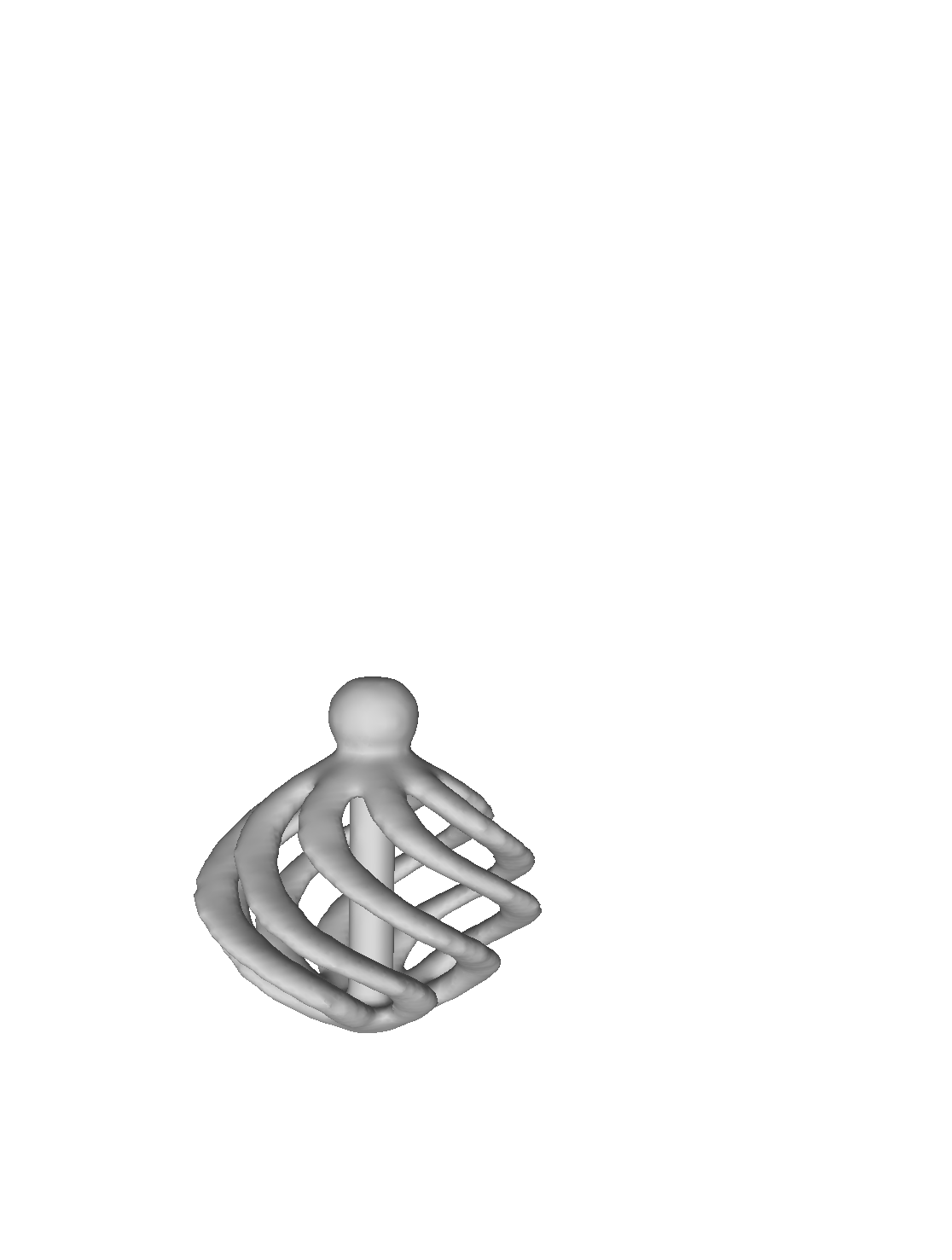}& \includegraphics[width=0.22\textwidth,clip,keepaspectratio,trim={2.2cm 2.5cm 5cm 8cm},clip]{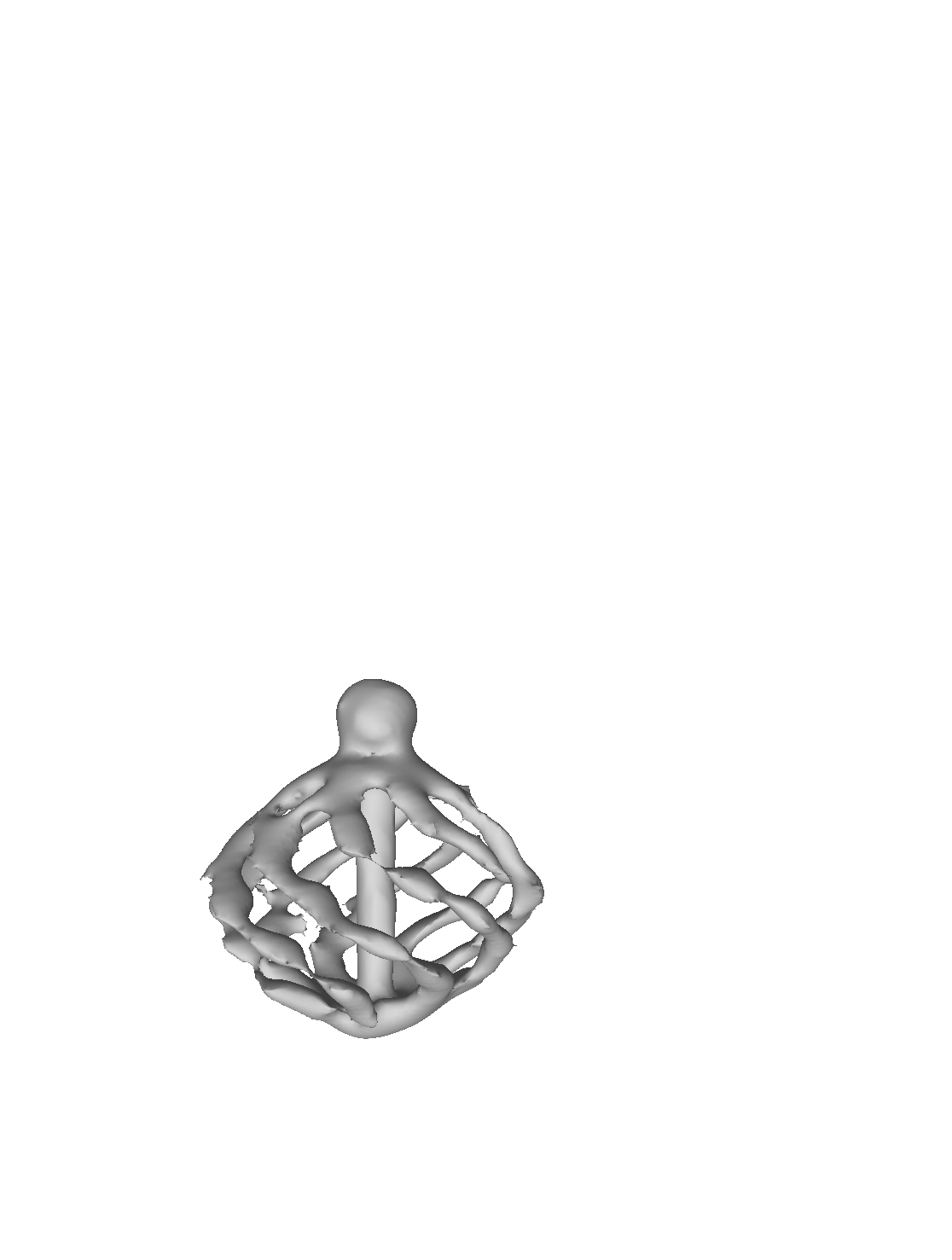}& \includegraphics[width=0.22\textwidth,clip,keepaspectratio,trim={2.2cm 2.5cm 5cm 8cm},clip]{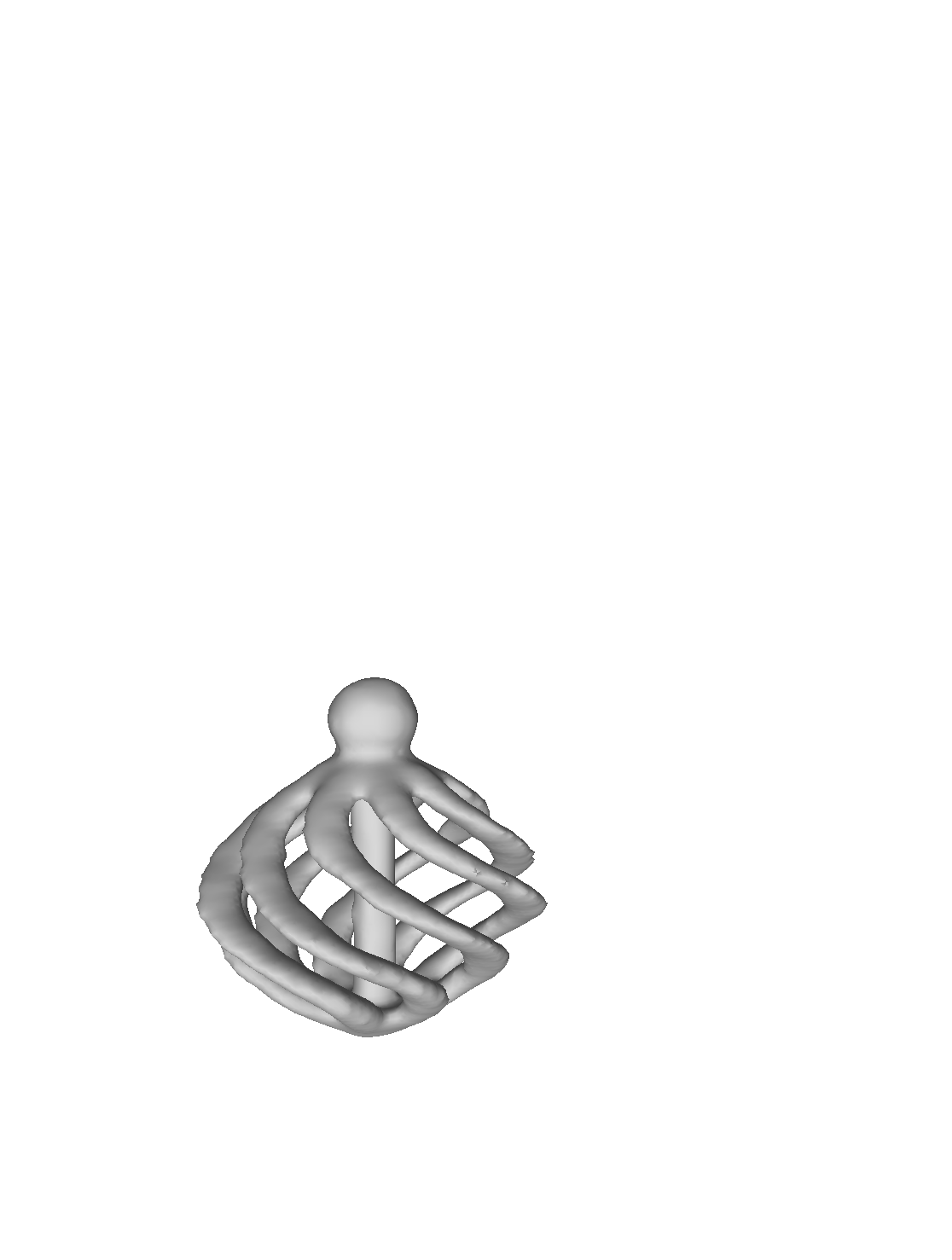}
\end{tabular}
 \caption{Joint reconstruction from tomographic and visual data. (a) Input object. Reconstructions based on (b) 800 dips, (c) 200 dips, and (d) 200 dips and 16 silhouettes.}
    \label{fig:dipVisFigure}
\end{figure}

Using the formulation in Eq. \eqref{jointMisfit}, we are able to use data from two different models to perform joint shape reconstruction. The first model that we use is the dip transform as in the previous experiments.
The second input source is a modified version of the SfS algorithm that was suggested in Section \ref{sec:new_SfS}.
Figure \ref{fig:dipVisFigure} shows that the use of the dip transform technique alone necessitates a large number of measurements (800) to obtain a good reconstruction, and achieves poor quality when fewer measurements (200) are used. The addition of the visual source significantly improves the reconstruction while reducing the number of the required dips. Thus, we conclude that the joint inversion algorithm is superior in terms of reconstruction accuracy. In addition, perhaps more importantly its use enables a considerable reduction in the number of data measurements needed, a favorable outcome that could promote such experiments which are typically expensive, difficult and/or dangerous to conduct (e.g., CT scans) when using conventional techniques.

\begin{figure}
    \centering
    \begin{subfigure}[b]{1.0\textwidth}
        \includegraphics[width=\textwidth]{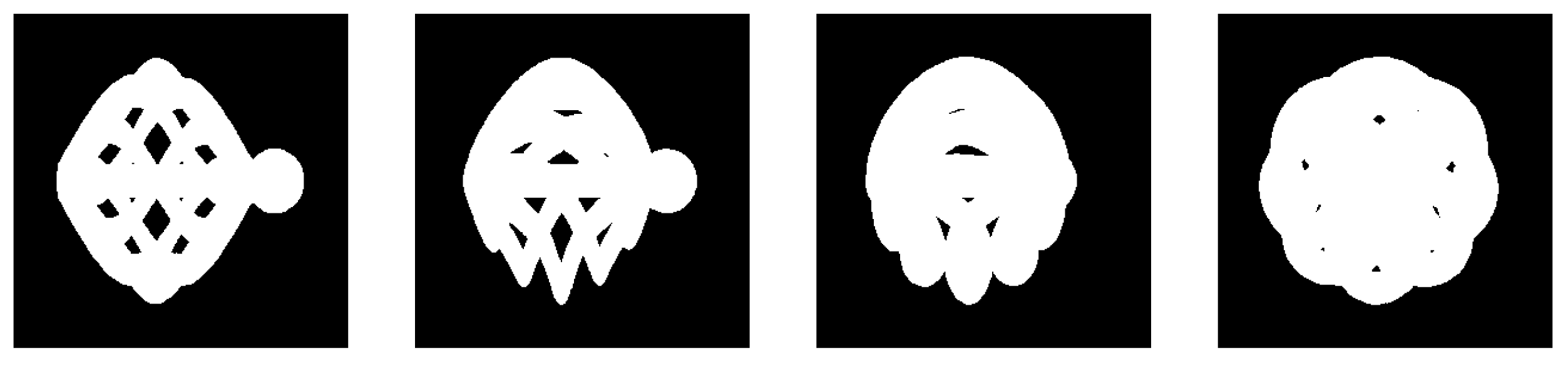}
    \end{subfigure}
\caption{Silhouettes of the object from different orientations obtained using our modified SfS approach described in \ref{sec:new_SfS}.}
\label{fig:SillhouettesIllustrations}
\end{figure}

\subsection{Shape reconstruction from Point Cloud} \label{sec:pointcloud}

\begin{figure}
    \centering
     \begin{subfigure}[b]{0.24\textwidth}
      \includegraphics[width=\textwidth,keepaspectratio,clip]{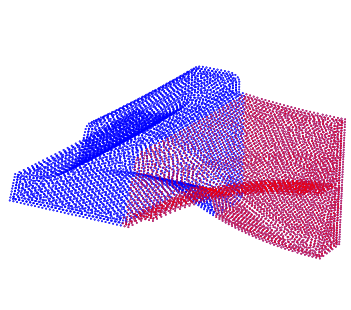}\caption{}\label{fig:PC_aligned}
    \end{subfigure}%
    \begin{subfigure}[b]{0.24\textwidth}
        \includegraphics[width=\textwidth,keepaspectratio,clip]{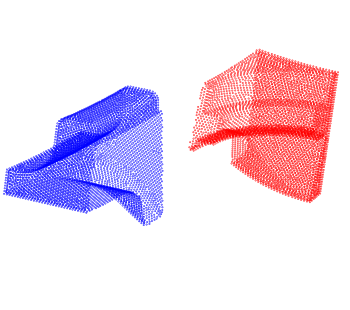}\caption{}\label{fig:PC}
    \end{subfigure}%
    \begin{subfigure}[b]{0.24\textwidth}
        \includegraphics[width=\textwidth,keepaspectratio,clip]{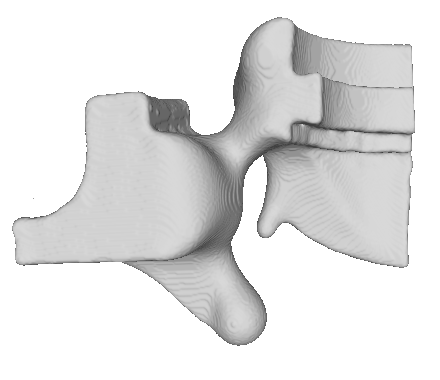}\caption{}\label{fig:reconPC}
    \end{subfigure}%
     \begin{subfigure}[b]{0.24\textwidth}
        \includegraphics[width=\textwidth,keepaspectratio,clip]{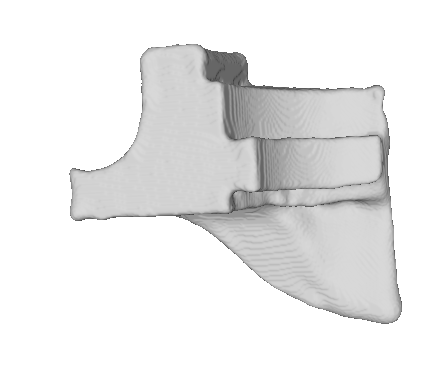}\caption{}\label{fig:reconPC_Vis}
    \end{subfigure}
\caption{Reconstruction from two point clouds. (a) The aligned point clouds. (b) The misaligned point clouds used as input. (c) A reconstruction from the misaligned point clouds only. (d) A joint reconstruction from the misaligned point clouds together with SfS.}
\label{fig:PointCloudResults}
\end{figure}

In this experiment we demonstrate our method for shape reconstruction from point clouds, as well as registration of point clouds. Often, two or more point clouds are given, possibly with some overlapping regions, and the goal is to align them such that they can be viewed as one point cloud for performing tasks like shape reconstruction.
To demonstrate the registration and reconstruction from point clouds we generate two non-overlapping point clouds from the shape in Fig. \ref{fig:fandisk}---see Fig. \ref{fig:PC_aligned}. Then we significantly move one of the point clouds as shown in Fig. \ref{fig:PC}, and try to reconstruct the shape. Since there is no overlap in the point distribution, methods like ICP often struggle to find the correct alignment. Fig. \ref{fig:reconPC} shows the a reconstruction using our mesh-free implementation. From an optimization perspective, this reconstruction is good even though it is misaligned---the data term in this experiment was reduced by two orders of magnitude in the optimization process. The drop-like artifacts that appear in the reconstruction are not visible in the misfit term involving the point cloud. We note that our method is able to recover the right alignment for some cases of initialization and choice of parameters, but that is not guaranteed by a low data term, and is not robust. To guarantee the correct reconstruction following a successful optimization, we add another data modality, SfS, which includes silhouettes of the complete shape. Fig \ref{fig:reconPC_Vis} shows a joint reconstruction from both the point clouds and 8 figures of silhouettes. This time, the point clouds are aligned by the optimization since it has the SfS to guide it towards the right shape, and recover the alignment parameters.

\section{Conclusions and Future Work}
In this paper, we proposed a general framework for robust and efficient 3D shape reconstruction using parametric level set methods with ellipsoidal radial basis functions. Using a compact PaLS representation, our reconstruction process estimates a relatively low number of parameters, and therefore, it requires a correspondingly small number of data measurements. Our framework has the ability to accurately estimate acquisition parameters as part of the reconstruction, and it is robust to noise that originates from inaccuracies in these parameters. Our framework also supports multi-modal reconstruction, and thus, it can be adapted to incorporate several models in a single shape estimation. The proposed method is capable of reducing the number of required measurements, which can generally be beneficial for models where experiments are expensive, difficult or dangerous to conduct (e.g., MRI  scans).

The compact analytical PaLS representation can be used to enhance various applications. For example, it is a natural approach to apply reconstruction from tomographic measurements under deformation \cite{zang2018space}, where the deformations can be treated analytically using the PaLS representations, and the number of measurements can be reduced.
The PaLS can also be considered as a regularization inside a deep neural network architecture, where the insights from this work can be incorporated to reduce the computational cost for 3D shape analysis and image segmentation.

\section{Appendix}\label{sec:Appendix}
\bigskip
\subsection{The Wendland's and Heaviside Functions}
\label{sec:heavyside}
The Wendland's compactly supported radial basis functions \cite{wendland1995piecewise} constitute a family of polynomial RBFs $\psi_i$ whose range is $[0,1]$ and zero outside $[0,1]$ (for $r>1$, $\psi_i(r) = 0$). A common choice to be used in $\mathbb{R}^2$ and $\mathbb{R}^3$ is one of the following:
\begin{eqnarray}
\psi_0(r) &=& (1-r)_+^2 \in C^{0}\\
\psi_1(r) &=& (1-r)_+^4(4r+1)\in C^{2}\\
\psi_2(r) &=& (1-r)_+^6\textstyle{\frac{1}{3}}(35r^2+4r+3)\in C^{4}\\
\psi_3(r) &=& (1-r)_+^6(32r^3 + 25r^2 + 8r + 1)\in C^{6}
\end{eqnarray}
where $()_+$ denotes the $max(x,0)$ function. In this work, we use $\psi_1$ because it is the simplest differentiable function and we did not see any change in our experiments when using functions of higher order.

The Heaviside function $\sigma$ smoothly transitions between 0 and 1. As shown in Fig. \ref{fig:heavyside} and mentioned in \cite{kadu2017parametric}, we wish to make the derivative of this function (the delta function) as flat as possible to render our Gauss-Newton Hessian better conditioned. We define our Heaviside function to be a smoothed linear transition from 0 to 1 using a piecewise polynomial function. More precisely, it is defined by a transition width $\delta$ and a smoothness width $\epsilon$ as follows:
\begin{equation}
\sigma_{\delta,\epsilon}(x) = \left\{
\begin{array}{cc}
0 & x < -\delta-\epsilon\\
\frac{1}{8\delta\epsilon}(x+\delta+\epsilon)^2 & -\delta-\epsilon \leq x < - \delta + \epsilon\\
\frac{1}{2\delta}(x+\delta) & -\delta+\epsilon \leq x <  \delta - \epsilon\\
-\frac{1}{8\delta\epsilon}(\delta+\epsilon-x)^2+1 & \delta-\epsilon \leq x < \delta + \epsilon\\
1 &  \delta + \epsilon < x
\end{array}\right.
\end{equation}

This function and its first derivative are continuous, and it is very similar to the function proposed in \cite{kadu2017parametric}, using a $sin$ function to define the transition smoothness. We prefer a polynomial approximation because of computation considerations. As default parameters throughout all the experiments in this paper, we choose $\delta=0.1,\epsilon=0.01$, and use the letter $\sigma$ to denote the function.

\subsection{Derivatives of the Ellipsoidal PaLS Function}\label{AppendixPaLS_Derivatives}
Here we show the derivatives of the ellipsoidal PaLS function, and the derivatives of the standard PaLS in \eqref{eq:pals} are defined in \cite{aghasi2011parametric}.
The $i$-th basis function in the sum of \eqref{enrichedPaLS} is defined by ${\alpha_i,B_i,\vec{\xi_i}}$. For each RBF (containing 10 parameters), we define its radius by:
\begin{equation}
r_i(\vec{x}) = \left\|\vec{x}-\vec{\xi}_i\right\|_{B_i}^\dag.
\end{equation}
Dropping the index $i$, the matrix $B$ is defined by 6 parameters due to symmetry:
\begin{eqnarray}\label{eq:symmB}
B = \left[\begin{array}{ccc}
B^{1} & B^2 & B^3\\
B^2 & B^4  & B^5\\
B^3 & B^5 & B^6\\
\end{array}\right].
\end{eqnarray}
We define the operator
\begin{eqnarray}\label{Pmatrix}
P = \left[\begin{array}{cccccc}
1 & 0 & 0 & 0 & 0 & 0\\
0 & 1  & 0 & 0 & 0 & 0\\
0 & 0  & 1 & 0 & 0 & 0\\
0 & 1  & 0 & 0 & 0 & 0\\
0 & 0  & 0 & 1 & 0 & 0\\
0 & 0  & 0 & 0 & 1 & 0\\
0 & 0  & 1 & 0 & 0 & 0\\
0 & 0  & 0 & 0 & 1 & 0\\
0 & 0  & 0 & 0 & 0 & 1\\
\end{array}\right]
\end{eqnarray}
to be the operator that takes the vector $[B^1,...,B^6]^\top$ and by multiplication transfers it into a vector of size 9 that corresponds to the symmetric matrix $B$ in \eqref{eq:symmB}.  With this notation, the derivatives of $r_i(\vec x)$ are computed by:
\begin{eqnarray}\label{eq:sens_r}
\frac{\partial r_i}{\partial \vec{\xi}_i} &=& -\frac{B_i\vec{z}_i}{r(\vec{x})}\in\mathbb{R}^3\\
\frac{\partial r_i}{\partial B_i} &=& \frac{1}{r(\vec{x})}P^\top vec(\vec{z}_i\vec{z}_i^\top)\in\mathbb{R}^6
\end{eqnarray}
where $\vec{z}_i = \vec{x}-\vec{\xi}_i$, and the operator $vec$ is the column-stack operator transferring a $3\times3$ matrix into a vector of size 9. Now we can easily define the derivatives of $u$, as the additional terms are just scalars:
\begin{eqnarray}
\frac{\partial u}{\partial \alpha_i}(\vec{x}) &=&  \sigma'(u(\vec{x})) \psi(r_i(\vec{x}))\\
\frac{\partial u}{\partial \vec{\xi}_i}(\vec{x}) &=& \alpha_i\sigma'(u(\vec{x})) \psi'(r_i(\vec{x}))\frac{\partial r_i}{\partial \vec{\xi}_i}\\
\frac{\partial u}{\partial B_i}(\vec{x}) &=& \alpha_i\sigma'(u(\vec{x})) \psi'(r_i(\vec{x}))\frac{\partial r_i}{\partial B_i}
\end{eqnarray}

\subsection{The derivatives of the rotated PaLS}\label{sec:PaLS_RotJacobians}
As noted before, the radial basis functions can be rotated analytically. Thus, instead of rotating $\bfu(\bfm)$ we can simply rotate each of the basis functions according to its rotation parameters. For the spherical RBFs, we have the rotation function in \eqref{eq:simple_mrot} for rotating the $i$-th basis function at angles $\theta_j,\phi_j$ and translation vector $\vec{b_j}$, which corresponds to the j-th measurements of the shape. The Jacobian of  \eqref{eq:simple_mrot} with respect to $(\alpha_i,\beta_i,\vec{\xi}_i)$ is equal to
\begin{equation}
\frac{\partial rot_j(\alpha_i,\beta_i,\vec{\xi}_i)}{\partial (\alpha_i,\beta_i,\vec{\xi}_i)} = \left[\begin{array}{ccc}
1 & 0 & \vec{0}^T\\
0 & 1  & \vec{0}^T\\
\vec{0} & \vec{0} & Q_j\\
\end{array}\right] \in\mathbb{R}^{5\times 5}.
\end{equation}

For the ellipsoidal RBFs, the rotation function is defined in \eqref{eq:rot_enriched} and its Jacobian is defined by
\begin{eqnarray}
\frac{\partial rot_j(\alpha_i,B_i,\vec{\xi}_i)}{\partial (\alpha_i,B_i,\vec{\xi}_i)} = \left[\begin{array}{ccc}
1 & 0_{1\times6} & 0_{1\times3}\\
0_{6\times1} & tril(Q_{j}\otimes Q_{j}P)_{6 \times 6}  & 0_{6\times3}\\
0_{3\times1} & 0_{3\times6} & Q_j\\
\end{array}\right]\in\mathbb{R}^{10\times 10},
\end{eqnarray}
where $P$ is the matrix defined in \eqref{Pmatrix} and $tril(A)$ is an operator that chooses the 6 out of 9 rows of $A$ corresponding to the indices of the lower-triangular elements of a $3 \times 3$ matrix. With this operator, the term $tril(Q_{j}\otimes Q_{j}P)\in\mathbb{R}^{6\times 6}$.

\subsection{Derivatives with respect to acquisition parameters}\label{appendixNoiseHandling}
To incorporate the rotation parameters in the optimization process, we need to compute the derivatives of $u$ according to them. This way, we are able to overcome uncertainty of the calibration in the data-acquisition process.
For spherical RBFs, we have to take the derivatives of \eqref{eq:simple_mrot} with respect to the additional 5 parameters in $(\theta_j,\phi_j,\vec{b}_j)$. We obtain:

\begin{eqnarray}
\frac{\partial rot(\alpha_i,\beta_i,\vec{\xi}_i,\theta_j,\phi_j,\vec{b}_j)}{\partial \theta_j} &=& [0,0,\frac{\partial Q_j}{\partial \theta_j}(\vec{x}-\vec{x}_{mid})]^T
 = [0,0,\vec{z}_{\theta_j}]^T \in \mathbb{R}^{5}\\
\frac{\partial rot(\alpha_i,\beta_i,\vec{\xi}_i,\theta_j,\phi_j,\vec{b}_j)}{\partial \phi_j} &=& [0,0,\frac{\partial Q_j}{\partial \phi_j}(\vec{x}-\vec{x}_{mid})]^T
 = [0,0,\vec{z}_{\phi_j}]^T\in \mathbb{R}^{5}\\
\frac{\partial rot(\alpha_i,\beta_i,\vec{\xi}_i,\theta_j,\phi_j,\vec{b}_j)}{\partial \vec{b}_j} &=& \begin{bmatrix} 0_{2\times3} \\ I_{3\times3} \end{bmatrix}\in\mathbb{R}^{5\times3}\label{eq:b_der}
\end{eqnarray}
where
\begin{equation}
\vec{z}_{\theta_j}=\frac{\partial Q_j}{\partial \theta_j}(\vec{x}-\vec{x}_{mid}),
\mbox{ and }
\vec{z}_{\phi_j}=\frac{\partial Q_j}{\partial \phi_j}(\vec{x}-\vec{x}_{mid}).
\end{equation}

Placing all the partial derivatives together in the Jacobian we obtain:
\begin{equation}\label{eq:noiseExtendedJacobian}
 \frac{\partial rot(\alpha_i,\beta_i,\vec{\xi}_i,\theta_j,\phi_j,\vec{b}_j)}{\partial
  (\alpha_i,\beta_i,\vec{\xi}_i,\theta_j,\phi_j,\vec{b}_j)} = \begin{bmatrix}
        I_{2\times2} &  0_{2 \times 3} & 0_{2 \times 1} & 0_{2 \times 1} &  0_{2 \times 3}\\
        0_{3 \times 2} & Q_{j} & \vec{z}_{\theta_j} & \vec{z}_{\phi_j} &I_{3 \times 3}
     \end{bmatrix}_{5 \times 10}
\end{equation}

For ellipsoidal RBFs, the derivatives of \eqref{eq:rot_enriched} include:

\begin{eqnarray}
\frac{\partial rot(\alpha_i,B_i,\vec{\xi}_i,\theta_j,\phi_j,\vec{b}_j)}{\partial (\theta_j,\phi_j)} =  \begin{bmatrix} 0 & 0\\\vec{T}_{\theta_j}& \vec{T}_{\phi_j}\\ \vec{z}_{\theta_j}& \vec{z}_{\phi_j}\end{bmatrix}\in\mathbb{R}^{10\times2},
\end{eqnarray}
where
\begin{eqnarray}
\vec{T}_{\theta_j} &=& tril\left(Q_{j}B_{i}\frac{\partial Q_{j}^{T}}{\partial \theta_{j}}+\frac{\partial Q_{j}}{\partial \theta_{j}}B_{i}Q_{j}^{T}\right)\in\mathbb{R}^{6}\\
\vec{T}_{\phi_j} &=& tril\left(Q_{j}B_{i}\frac{\partial Q_{j}^{T}}{\partial \phi_{j}}+\frac{\partial Q_{j}}{\partial \phi_{j}}B_{i}Q_{j}^{T}\right)\in\mathbb{R}^{6},
\end{eqnarray}
and the derivative with respect to $\vec{b}_j$ is similar to \eqref{eq:b_der}. Altogether, we obtain the Jacobian:
\begin{equation}\label{eq:noiseExtendedJacobian_10basedRBF}
 \frac{\partial rot(\alpha_i,B_i,\vec{\xi}_i,\theta_j,\phi_j,\vec{b}_j)}{\partial
  (\alpha_i,B_i,\vec{\xi}_i,\theta_j,\phi_j,\vec{b}_j)} = \left[\begin{array}{cccccc}
1 & 0_{1\times6} & 0_{1\times3} & 0 & 0 & 0_{1\times3}\\
\vec{0} & tril(Q_{j}\otimes Q_{j}P)_{6 \times 6} & 0_{6\times3} & \vec{T}_{\theta_j} & \vec{T}_{\phi_j} & 0_{6\times3}  \\
\vec{0} & 0_{3 \times 6} & Q_j &  \vec{z}_{\theta_j} &  \vec{z}_{\phi_j} & I_{3 \times 3}\\
\end{array}\right]_{10 \times 15}.
\end{equation}

\footnotesize
\bibliographystyle{siam} 
\bibliography{reconstructionbib}

\begin{thebibliography}{10}

\bibitem{aberman2017dip}
{\sc K.~Aberman, O.~Katzir, Q.~Zhou, Z.~Luo, A.~Sharf, C.~Greif, B.~Chen, and
  D.~Cohen-Or}, {\em Dip transform for 3d shape reconstruction}, ACM
  Transactions on Graphics (TOG), 36 (2017), p.~79.

\bibitem{aghasi2011parametric}
{\sc A.~Aghasi, M.~Kilmer, and E.~L. Miller}, {\em Parametric level set methods
  for inverse problems}, SIAM Journal on Imaging Sciences, 4 (2011),
  pp.~618--650.

\bibitem{aghasi2013sparse}
{\sc A.~Aghasi and J.~Romberg}, {\em Sparse shape reconstruction}, SIAM Journal
  on Imaging Sciences, 6 (2013), pp.~2075--2108.

\bibitem{avron2010L1}
{\sc H.~Avron, A.~Sharf, C.~Greif, and D.~Cohen-Or}, {\em $\ell_1$-sparse
  reconstruction of sharp point set surfaces}, ACM Transactions on Graphics
  (TOG), 29 (2010), p.~135.

\bibitem{basu2000uniqueness}
{\sc S.~Basu and Y.~Bresler}, {\em Uniqueness of tomography with unknown view
  angles}, IEEE Transactions on Image Processing, 9 (2000), pp.~1094--1106.

\bibitem{berger2014state}
{\sc M.~Berger, A.~Tagliasacchi, L.~Seversky, P.~Alliez, J.~Levine, A.~Sharf,
  and C.~Silva}, {\em State of the art in surface reconstruction from point
  clouds}, EUROGRAPHICS star reports, 1 (2014), pp.~161--185.

\bibitem{berger2017survey}
{\sc M.~Berger, A.~Tagliasacchi, L.~M. Seversky, P.~Alliez, G.~Guennebaud,
  J.~A. Levine, A.~Sharf, and C.~T. Silva}, {\em A survey of surface
  reconstruction from point clouds}, in Computer Graphics Forum, vol.~36, Wiley
  Online Library, 2017, pp.~301--329.

\bibitem{bermano2011online}
{\sc A.~Bermano, A.~Vaxman, and C.~Gotsman}, {\em Online reconstruction of 3d
  objects from arbitrary cross-sections}, ACM Transactions on Graphics (TOG),
  30 (2011), p.~113.

\bibitem{besl1992method}
{\sc P.~J. Besl and N.~D. McKay}, {\em Method for registration of 3-d shapes},
  in Sensor fusion IV: control paradigms and data structures, vol.~1611,
  International Society for Optics and Photonics, 1992, pp.~586--606.

\bibitem{Julia}
{\sc J.~Bezanson, A.~Edelman, S.~Karpinski, and V.~B. Shah}, {\em Julia: A
  fresh approach to numerical computing}, SIAM Review, 59 (2017), pp.~65--98.

\bibitem{bleichrodt2013alignment}
{\sc F.~Bleichrodt, J.~Sijbers, J.~de~Beenhouwer, and K.~J. Batenburg}, {\em An
  alignment method for fan beam tomography}, in Proceedings of 1st
  International Conference on Tomography of Materials and Structures, 2013,
  pp.~103--106.

\bibitem{boumal2014manopt}
{\sc N.~Boumal, B.~Mishra, P.-A. Absil, and R.~Sepulchre}, {\em Manopt, a
  matlab toolbox for optimization on manifolds}, The Journal of Machine
  Learning Research, 15 (2014), pp.~1455--1459.

\bibitem{bretin2017volume}
{\sc E.~Bretin, F.~Dayrens, and S.~Masnou}, {\em Volume reconstruction from
  slices}, SIAM Journal on Imaging Sciences, 10 (2017), pp.~2326--2358.

\bibitem{carr2001reconstruction}
{\sc J.~C. Carr, R.~K. Beatson, J.~B. Cherrie, T.~J. Mitchell, W.~R. Fright,
  B.~C. McCallum, and T.~R. Evans}, {\em Reconstruction and representation of
  3d objects with radial basis functions}, in Proceedings of the 28th annual
  conference on Computer graphics and interactive techniques, ACM, 2001,
  pp.~67--76.

\bibitem{chen2018deep}
{\sc W.~Chen, X.~Han, G.~Li, C.~Chen, J.~Xing, Y.~Zhao, and H.~Li}, {\em Deep
  rbfnet: Point cloud feature learning using radial basis functions}, arXiv
  preprint arXiv:1812.04302,  (2018).

\bibitem{cremers2011multiview}
{\sc D.~Cremers and K.~Kolev}, {\em Multiview stereo and silhouette consistency
  via convex functionals over convex domains}, IEEE Trans. on Pattern Analysis
  and Machine Intelligence, 33 (2011), pp.~1161--1174.

\bibitem{cuomo2013surface}
{\sc S.~Cuomo, A.~Galletti, G.~Giunta, and A.~Starace}, {\em Surface
  reconstruction from scattered point via rbf interpolation on gpu}, in
  Computer Science and Information Systems (FedCSIS), 2013 Federated Conference
  on, IEEE, 2013, pp.~433--440.

\bibitem{de2015nonlinear}
{\sc E.~De~Sturler, S.~Gugercin, M.~E. Kilmer, S.~Chaturantabut, C.~Beattie,
  and M.~O'Connell}, {\em Nonlinear parametric inversion using interpolatory
  model reduction}, SIAM Journal on Scientific Computing, 37 (2015),
  pp.~B495--B517.

\bibitem{fung2018uncertainty}
{\sc S.~W. Fung and L.~Ruthotto}, {\em An uncertainty-weighted asynchronous
  admm method for parallel pde parameter estimation}, arXiv preprint
  arXiv:1806.00192,  (2018).

\bibitem{haber2014computational}
{\sc E.~Haber}, {\em Computational methods in geophysical electromagnetics},
  vol.~1, SIAM, 2014.

\bibitem{houben2011refinement}
{\sc L.~Houben and M.~B. Sadan}, {\em Refinement procedure for the image
  alignment in high-resolution electron tomography}, Ultramicroscopy, 111
  (2011), pp.~1512--1520.

\bibitem{kadu2017parametric}
{\sc A.~Kadu, T.~van Leeuwen, and K.~J. Batenburg}, {\em A parametric level-set
  method for partially discrete tomography}, in International Conference on
  Discrete Geometry for Computer Imagery, Springer, 2017, pp.~122--134.

\bibitem{kadu2017salt}
{\sc A.~Kadu, T.~van Leeuwen, and W.~A. Mulder}, {\em Salt reconstruction in
  full-waveform inversion with a parametric level-set method}, IEEE Trans. on
  Computational Imaging, 3 (2017), pp.~305--315.

\bibitem{kainz2015fast}
{\sc B.~Kainz, M.~Steinberger, W.~Wein, M.~Kuklisova-Murgasova,
  C.~Malamateniou, K.~Keraudren, T.~Torsney-Weir, M.~Rutherford, P.~Aljabar,
  J.~V. Hajnal, et~al.}, {\em Fast volume reconstruction from motion corrupted
  stacks of 2d slices}, IEEE transactions on medical imaging, 34 (2015),
  pp.~1901--1913.

\bibitem{kim2019three}
{\sc J.~Kim and C.-O. Lee}, {\em Three-dimensional volume reconstruction using
  two-dimensional parallel slices}, SIAM Journal on Imaging Sciences, 12
  (2019), pp.~1--27.

\bibitem{kolev2012fast}
{\sc K.~Kolev, T.~Brox, and D.~Cremers}, {\em Fast joint estimation of
  silhouettes and dense 3d geometry from multiple images}, IEEE Trans. on
  Pattern Analysis and Machine Intelligence, 34 (2012), pp.~493--505.

\bibitem{kolev2009continuous}
{\sc K.~Kolev, M.~Klodt, T.~Brox, and D.~Cremers}, {\em Continuous global
  optimization in multiview 3d reconstruction}, International Journal of
  Computer Vision, 84 (2009), pp.~80--96.

\bibitem{larusson2013parametric}
{\sc F.~Larusson, P.~G. Anderson, E.~Rosenberg, M.~E. Kilmer, A.~Sassaroli,
  S.~Fantini, and E.~L. Miller}, {\em Parametric estimation of 3d tubular
  structures for diffuse optical tomography}, Biomedical optics express, 4
  (2013), pp.~271--286.

\bibitem{laurentini1994visual}
{\sc A.~Laurentini}, {\em The visual hull concept for silhouette-based image
  understanding}, IEEE Transactions on pattern analysis and machine
  intelligence, 16 (1994), pp.~150--162.

\bibitem{liu2008surface}
{\sc L.~Liu, C.~Bajaj, J.~O. Deasy, D.~A. Low, and T.~Ju}, {\em Surface
  reconstruction from non-parallel curve networks}, Computer Graphics Forum, 27
  (2008), pp.~155--163.

\bibitem{mcmillan20153d}
{\sc M.~S. McMillan, C.~Schwarzbach, E.~Haber, and D.~W. Oldenburg}, {\em 3d
  parametric hybrid inversion of time-domain airborne electromagnetic data},
  Geophysics, 80 (2015), pp.~K25--K36.

\bibitem{metivier2017full}
{\sc L.~Metivier, R.~Brossier, S.~Operto, and J.~Virieux}, {\em Full waveform
  inversion and the truncated newton method}, SIAM Review, 59 (2017),
  pp.~153--195.

\bibitem{ohtake2003multi}
{\sc Y.~Ohtake, A.~Belyaev, and H.-P. Seidel}, {\em A multi-scale approach to
  3d scattered data interpolation with compactly supported basis functions}, in
  Shape Modeling International, 2003, IEEE, 2003, pp.~153--161.

\bibitem{parkinson2012automatic}
{\sc D.~Y. Parkinson, C.~Knoechel, C.~Yang, C.~A. Larabell, and M.~A. Le~Gros},
  {\em Automatic alignment and reconstruction of images for soft x-ray
  tomography}, Journal of structural biology, 177 (2012), pp.~259--266.

\bibitem{qi2017pointnet++}
{\sc C.~R. Qi, L.~Yi, H.~Su, and L.~J. Guibas}, {\em Pointnet++: Deep
  hierarchical feature learning on point sets in a metric space}, in Advances
  in Neural Information Processing Systems, 2017, pp.~5099--5108.

\bibitem{rudin1992nonlinear}
{\sc L.~I. Rudin, S.~Osher, and E.~Fatemi}, {\em Nonlinear total variation
  based noise removal algorithms}, Physica D: Nonlinear Phenomena, 60 (1992),
  pp.~259--268.

\bibitem{rusinkiewicz2001efficient}
{\sc S.~Rusinkiewicz and M.~Levoy}, {\em Efficient variants of the icp
  algorithm.}, in 3dim, vol.~1, 2001, pp.~145--152.

\bibitem{jInv17}
{\sc L.~Ruthotto, E.~Treister, and E.~Haber}, {\em {jInv} -- a flexible {Julia}
  package for {PDE} parameter estimation}, SIAM J. Sci. Comput., 39 (2017),
  pp.~S702–--S722.

\bibitem{santosa1996level}
{\sc F.~Santosa}, {\em A level-set approach for inverse problems involving
  obstacles fadil santosa}, ESAIM: Control, Optimisation and Calculus of
  Variations, 1 (1996), pp.~17--33.

\bibitem{sethian1999level}
{\sc J.~A. Sethian}, {\em Level set methods and fast marching methods: evolving
  interfaces in computational geometry, fluid mechanics, computer vision, and
  materials science}, vol.~3, Cambridge university press, 1999.

\bibitem{singer2011three}
{\sc A.~Singer and Y.~Shkolnisky}, {\em Three-dimensional structure
  determination from common lines in cryo-em by eigenvectors and semidefinite
  programming}, SIAM journal on imaging sciences, 4 (2011), pp.~543--572.

\bibitem{takimoto20163d}
{\sc R.~Y. Takimoto, M.~d. S.~G. Tsuzuki, R.~Vogelaar, T.~de~Castro~Martins,
  A.~K. Sato, Y.~Iwao, T.~Gotoh, and S.~Kagei}, {\em 3d reconstruction and
  multiple point cloud registration using a low precision rgb-d sensor},
  Mechatronics, 35 (2016), pp.~11--22.

\bibitem{tam2012registration}
{\sc G.~K. Tam, Z.-Q. Cheng, Y.-K. Lai, F.~C. Langbein, Y.~Liu, D.~Marshall,
  R.~R. Martin, X.-F. Sun, and P.~L. Rosin}, {\em Registration of 3d point
  clouds and meshes: a survey from rigid to nonrigid}, IEEE transactions on
  visualization and computer graphics, 19 (2012), pp.~1199--1217.

\bibitem{JointEikFWI17}
{\sc E.~Treister and E.~Haber}, {\em Full waveform inversion guided by travel
  time tomography}, SIAM J. Sci. Comput., 39 (2017), pp.~S587--–S609.

\bibitem{tsai2017indoor}
{\sc C.-Y. Tsai and C.-H. Huang}, {\em Indoor scene point cloud registration
  algorithm based on rgb-d camera calibration}, Sensors, 17 (2017), p.~1874.

\bibitem{van2018automatic}
{\sc T.~van Leeuwen, S.~Maretzke, and K.~J. Batenburg}, {\em Automatic
  alignment for three-dimensional tomographic reconstruction}, Inverse
  Problems, 34 (2018), p.~024004.

\bibitem{vese2002multiphase}
{\sc L.~A. Vese and T.~F. Chan}, {\em A multiphase level set framework for
  image segmentation using the mumford and shah model}, International journal
  of computer vision, 50 (2002), pp.~271--293.

\bibitem{vogel2002computational}
{\sc C.~R. Vogel}, {\em Computational methods for inverse problems}, vol.~23,
  SIAM, Philadelphia, 2002.

\bibitem{wendland1995piecewise}
{\sc H.~Wendland}, {\em Piecewise polynomial, positive definite and compactly
  supported radial functions of minimal degree}, Advances in computational
  Mathematics, 4 (1995), pp.~389--396.

\bibitem{whitaker1998level}
{\sc R.~T. Whitaker}, {\em A level-set approach to 3d reconstruction from range
  data}, International journal of computer vision, 29 (1998), pp.~203--231.

\bibitem{yang2005unified}
{\sc C.~Yang, E.~G. Ng, and P.~A. Penczek}, {\em Unified 3-d structure and
  projection orientation refinement using quasi-newton algorithm}, Journal of
  structural biology, 149 (2005), pp.~53--64.

\bibitem{yin2016full}
{\sc K.~Yin, H.~Huang, P.~Long, A.~Gaissinski, M.~Gong, and A.~Sharf}, {\em
  Full 3d plant reconstruction via intrusive acquisition}, Computer Graphics
  Forum, 35 (2016), pp.~272--284.

\bibitem{zang2018space}
{\sc G.~Zang, R.~Idouchi, R.~Tao, G.~Lubineau, P.~Wonka, and W.~Heidrich}, {\em
  Space-time tomography for continuously deforming objects}, ACM Transactions
  on Graphics (TOG), 37 (2018), p.~100.

\bibitem{zheng20174d}
{\sc Q.~Zheng, X.~Fan, M.~Gong, A.~Sharf, O.~Deussen, and H.~Huang}, {\em 4d
  reconstruction of blooming flowers}, Computer Graphics Forum, 36 (2017),
  pp.~405--417.

\bibitem{zou2015topology}
{\sc M.~Zou, M.~Holloway, N.~Carr, and T.~Ju}, {\em Topology-constrained
  surface reconstruction from cross-sections}, ACM Transactions on Graphics
  (TOG), 34 (2015), p.~128.

\end{thebibliography}
\pagenumbering{gobble}

\end{document}


\begin{remark}
To make the basis function even more expressive, we enrich each basis function to represent an ellipsoid in a certain orientation. This is done by replacing $\beta$ by a lower triangular matrix $L\in\mathbb{R}^{3\times 3}$:
\begin{equation}
u(\vec{x},\alpha_i,L_i,\vec{\xi}_i) = \sigma_{\delta} \left(\sum_i{\alpha_i\psi\left(\left\|L_i^T\left(\vec{x}-\vec{\xi}_i\right)\right\|_2^\dag\right)}\right).
\end{equation}
Using the six parameters for $L_i$, we represent all the symmetric positive definite $3\times3$ matrices by multiplying $L_iL_i^T$ (in this case, $L_i$ is the Cholesky factor of any SPD matrix).

To compute the sensitivities, we first denote the radius
$$
r(\vec{x}) = \left\|L_i^\top\left(\vec{x}-\vec{\xi}_i\right)\right\|_2^\dag,
$$
and compute its derivatives:
\begin{eqnarray}\label{eq:sens_r2}
\frac{\partial r}{\partial \vec{\xi}_i} &=& -\frac{L_iL_i^\top\vec{z}_i}{r(\vec{x})}\\
\frac{\partial r}{\partial L_i} &=& \frac{1}{r(\vec{x})}\mbox{tril}(\vec{z}_i\vec{z}_i^\top L_i)
\end{eqnarray}
where $z_i = \vec{x}-\vec{\xi}_i$, and $\mbox{tril}()$ denotes the lower triangular part of any matrix.
Then:
\begin{eqnarray}
\frac{\partial u}{\partial \alpha_i}(\vec{x}) &=&  \sigma'(u(\vec{x})) \psi(r(\vec{x}))\\
\frac{\partial u}{\partial \vec{\xi}_i}(\vec{x}) &=& \alpha_i\sigma'(u(\vec{x})) \psi'(r(\vec{x}))\frac{\partial r}{\partial \vec{\xi}_i}\\
\frac{\partial u}{\partial L_i}(\vec{x}) &=& \alpha_i\sigma'(u(\vec{x})) \psi'(r(\vec{x}))\frac{\partial r}{\partial L_i}
\end{eqnarray}
\end{remark}

\section{enriched rbfs}
$$
\mbox{vec}(R^{-\top}\delta A_iR^{-1}) = R^{-\top}\otimes R^{-\top}\mbox{vec}(\delta A_i)
$$

$$
\mbox{vec}(\delta A_i) = P\mbox{trilvec}(\delta A_i)
$$
$$
\mbox{trilvec}(R^{-\top}\delta A_iR^{-1}) = \mbox{trilvec}(R^{-\top}\otimes R^{-\top}P\mbox{trilvec}(\delta A_i))
$$
\begin{eqnarray}\varphi
R(\theta+\delta\theta,\varphi+\delta\varphi) \approx R+\delta\theta R'_\theta + \delta\varphi R_\varphi'
\end{eqnarray}

\section{rotating parameters}
\begin{equation}\label{urot}
\bfu_j(\vec{x},\bfm) = \bfu(T_j^{-1}(\vec{x}),\bfm).
\end{equation}
\\
Where the inverse transformation is:
\begin{equation}\label{inverseTrans}
T_j^{-1}(\vec{x}) = Q(\theta_j,\phi_j)^{-1}(\vec{x} - \vec{x}_{mid} - \vec{b}) + \vec{x}_{mid}
\end{equation}\\

\subsection{Rotation for basic RBFs}
In this subsection we develop the rotation for basic RBF parameter. Recall the PaLS representation:
\begin{equation}\label{recallPaLS}
u(\vec{x},\alpha_i,\beta_i,\vec{\xi}_i) = \sigma_{\delta} \left(\sum_i{\alpha_i\psi\left(\left\|\beta_i\left(\vec{x}-\vec{\xi}_i\right)\right\|_2^\dag\right)}\right)
\end{equation}
Plugging \ref{rotatedShape} into \ref{recallPaLS} we obtain:
\begin{eqnarray}
u(T^{-1}(\vec{x}),\bfm) &=& \sigma \left(\sum_i{\alpha_i\psi\left(\left\|\beta_i\left(T^{-1}(\vec{x})-\vec{\xi}_i\right)\right\|_2^\dag\right)}\right)\\
&=& \sigma \left(\sum_i{\alpha_i\psi\left(\left\|\beta_i\left(Q(\theta_j,\phi_j)^{-1}(\vec{x} - \vec{x}_{mid} - \vec{b}_j) + \vec{x}_{mid}-\vec{\xi}_i\right)\right\|_2^\dag\right)}\right)\\
&=& \sigma \left(\sum_i{\alpha_i\psi\left(\left\|\beta_i\left(\vec{x} - \vec{x}_{mid} - \vec{b}_j + Q(\theta_j,\phi_j)(\vec{x}_{mid}-\vec{\xi}_i)\right)\right\|_2^\dag\right)}\right)\\
&=& \sigma \left(\sum_i{\alpha_i\psi\left(\left\|\beta_i\left(\vec{x}-T_j(\vec{\xi}_i)\right)\right\|_2^\dag\right)}\right)
\end{eqnarray}
The third equality holds because $Q$ is an orthogonal matrix which does not change the $\ell_2$ norm of a vector. Note, $rot_j(\bfm)$ only rotates the center of the basis function.\\

At last, we calculate the Jacobian for the rotation function, as we will use it later in the optimization task.
\begin{eqnarray}
\nabla rot_j(\{\alpha_i,\beta_i,\vec{\xi}_i\}) = \left[\begin{array}{ccc}
1 & 0 & \vec{0}^T\\
0 & 1  & \vec{0}^T\\
\vec{0} & \vec{0} & Q\\
\end{array}\right] 
\end{eqnarray}

\subsection{Rotation for Enriched RBFs - Ellipsoids}
To further enhance the accuracy and expressiveness of the suggested framework we incorporate the enriched RBF from \ref{enriched} and develop the mathematical machinery to perform rotations their parameters.

Similarly to the previous subsection, we formulate the rotation of our proposed enriched RBFs. This,however, is slightly more complicated and requires some tricks. 

\begin{eqnarray}
& &u(T^{-1}(\vec{x}),\bfm)\\
&=& \sigma \left(\sum_i{\alpha_i\psi\left(\left\|\left(T^{-1}(\vec{x})-\vec{\xi}_i\right)\right\|_{B_i}^\dag\right)}\right)\\
&=& \sigma \left(\sum_i{\alpha_i\psi\left(\left\|Q_{j}^{-1} Q_{j}\left(T^{-1}(\vec{x})-\vec{\xi}_i\right)\right\|_{B_{i}}^\dag\right)}\right)\\
&=& \sigma \left(\sum_i{\alpha_i\psi\left(\left\|B_{i}^{\frac{1}{2}} Q_{j}^{-1} Q_{j}\left(T^{-1}(\vec{x})-\vec{\xi}_i\right)\right\|_{2}^\dag\right)}\right)\\
&=& \sigma \left(\sum_i{\alpha_i\psi\left(\left\|B_{i}^{\frac{1}{2}} Q_{j}^{-1} Q_{j}\left( Q(\theta_j,\phi_j)^{-1}(\vec{x} - \vec{x}_{mid} - \vec{b}_j) + \vec{x}_{mid}-\vec{\xi}_i \right)\right\|_{2}^\dag\right)}\right)\\
&=& \sigma \left(\sum_i{\alpha_i\psi\left(\left\|B_{i}^{\frac{1}{2}} Q_{j}^{-1} \left( (\vec{x} - \vec{x}_{mid} - \vec{b}_j) + Q(\theta_j,\phi_j)(\vec{x}_{mid}-\vec{\xi}_i) \right)\right\|_{2}^\dag\right)}\right)\\
&=& \sigma \left(\sum_i{\alpha_i\psi\left(\left\|B_{i}^{\frac{1}{2}} Q_{j}^{-1} \left( (\vec{x} - \vec{x}_{mid} - \vec{b}_j) + T_{j}(\vec{\xi}_i) \right)\right\|_{2}^\dag\right)}\right)
\end{eqnarray}

And the Jacobian is as follows:

\begin{eqnarray}
\nabla rot_j(\{\alpha_i,\beta_i,\vec{\xi}_i\}) = \left[\begin{array}{ccc}
1 & 0 & \vec{0}^T\\
0 & tril(Q_{j}^{-T}\otimes Q_{j}^{-T}P)_{6 \times 6}  & \vec{0}^T\\
\vec{0} & \vec{0} & Q\\
\end{array}\right]_{10 \times 10}
\end{eqnarray}

TODO: put in $Q_{j}^{-T}\otimes Q_{j}^{-T} P$ and take only the rows in $trilA()$.
TODO later against code: use $Q_j$ instead of $Q^{-T}$

\section{Handling noise in basic RBF}
For the basic case of spheres - using Eq.20 and Eq.21 we get:
\[
\frac{\partial \bfm_j^i}{\partial \theta_j} = [0,0,\frac{\partial \boldsymbol{Q_j}}{\partial \theta_j}(\vec{x}-\vec{x_{mid}})]
 = [0,0,Q_{t}z]
\]

\[
\frac{\partial \bfm_j^i}{\partial \phi_j} = [0,0,\frac{\partial \boldsymbol{Q_j}}{\partial \phi_j}(\vec{x}-\vec{x_{mid}})]
= [0,0,Q_{p}z]
\]

\[
\frac{\partial \bfm_j^i}{\partial \bfb_1} = [0,0,1,0,0]
\]
\[
\frac{\partial \bfm_j^i}{\partial \bfb_2} = [0,0,0,1,0]
\]
\[
\frac{\partial \bfm_j^i}{\partial \bfb_3} = [0,0,0,0,1]
\]
\\
Placing the partial derivatives in the Jacobian we obtain: 
\[ \boldsymbol{J}^{m_j^i} = \begin{bmatrix}
        1 & 0 & 0 & 0 & 0 & 0 & 0 & 0 & 0 & 0           \\ 
        0 & 1 & 0 & 0 & 0 & 0 & 0 & 0 & 0 & 0 \\
        0 & 0 & q_{1,1} & q_{1,2} & q_{1,3} & (Q_{t}z)_1 & 			(Q_{p}z)_1 & 1 & 0 & 0 \\
        0 & 0 & q_{2,1} & q_{2,2} & q_{2,3} & (Q_{t}z)_2 & 			(Q_{p}z)_2 & 0 & 1 & 0 \\
        0 & 0 & q_{3,1} & q_{3,2} & q_{3,3} & (Q_{t}z)_3 & 			(Q_{p}z)_3 & 0 & 0 & 1
     \end{bmatrix}_{5 \times 10}
\]

\section{Handling noise for enriched RBF}

\[
\frac{\partial \bfm_j}{\partial \theta_j} = [0,Q_{j}^{-T}B_{i}\frac{\partial Q_{j}^{-1}}{\partial \theta_{j}}+\frac{\partial Q_{j}^{-T}}{\partial \theta_{j}}B_{i}Q_{j}^{-1},Q_{t}z]
\]

\[
\frac{\partial \bfm_j}{\partial \phi_j} =  [0,Q_{j}^{-T}B_{i}\frac{\partial Q_{j}^{-1}}{\partial \phi_{j}}+\frac{\partial Q_{j}^{-T}}{\partial \phi_{j}}B_{i}Q_{j}^{-1},Q_{p}z]
\]

\[
\frac{\partial \bfm_j}{\partial \bfb_1} = [0,\vec 0,1]
\]

\[
\frac{\partial \bfm_j}{\partial \bfb_2} = [0,\vec 0,1]
\]

\[
\frac{\partial \bfm_j}{\partial \bfb_3} = [0,\vec 0,1]
\]

Thus, we obtain:

\begin{eqnarray}
J_{m_{extended}^{rot_j}} = \left[\begin{array}{cccccc}
1 & 0 & \vec{0}^T & \vec{0}^T & \vec{0}^T & \vec{0}^T\\
0 & tril(Q_{j}^{-T}\otimes Q_{j}^{-T}P)_{6 \times 6} & \vec{0}^T & \vec{T_t}_{3 \times 1} & \vec{T_p}_{3 \times 1} & \boldsymbol{0}  \\
\vec{0} & \vec{0}^T & Q & Q_t z_{3 \times 1} & Q_p z_{3 \times 1} & \boldsymbol{I}_{3 \times 3}\\
\end{array}\right]_{10 \times 15}
\end{eqnarray}

\subsection{rotation for ellipsoids}
\begin{eqnarray}
& &u(T^{-1}(\vec{x}),\bfm)\\
&=& \sigma \left(\sum_i{\alpha_i\psi\left(\left\|\left(T^{-1}(\vec{x})-\vec{\xi}_i\right)\right\|_{B_i}^\dag\right)}\right)\\
&=& \sigma \left(\sum_i{\alpha_i\psi\left(\left\|Q_{j}^{-1} Q_{j}\left(T^{-1}(\vec{x})-\vec{\xi}_i\right)\right\|_{B_{i}}^\dag\right)}\right)\\
&=& \sigma \left(\sum_i{\alpha_i\psi\left(\left\| Q_{j}^{-1} Q_{j}\left( Q(\theta_j,\phi_j)^{-1}(\vec{x} - \vec{x}_{mid} - \vec{b}_j) + \vec{x}_{mid}-\vec{\xi}_i \right)\right\|_{B_i}^\dag\right)}\right)\\
&=& \sigma \left(\sum_i{\alpha_i\psi\left(\left\| Q_{j}^{-1} \left( (\vec{x} - \vec{x}_{mid} - \vec{b}_j) + Q(\theta_j,\phi_j)(\vec{x}_{mid}-\vec{\xi}_i) \right)\right\|_{B_i}^\dag\right)}\right)\\
&=& \sigma \left(\sum_i{\alpha_i\psi\left(\left\| Q_{j}^{-1} \left( (\vec{x} - \vec{x}_{mid} - \vec{b}_j) + T_{j}(\vec{\xi}_i) \right)\right\|_{B_i}^\dag\right)}\right)\\
&=& \sigma \left(\sum_i{\alpha_i\psi\left(\left\|\left( (\vec{x} - \vec{x}_{mid} - \vec{b}_j) + T_{j}(\vec{\xi}_i) \right)\right\|_{Q_j^{-T}B_iQ_j^{-1}}^\dag\right)}\right)
\end{eqnarray}